\documentclass[nologo,11pt,a4paper,hidelinks]{ETHpaper}
\usepackage{graphicx, amsmath, amssymb,color,wasysym}
\usepackage[numbers]{natbib}
\usepackage{subcaption}
\usepackage{booktabs}
\usepackage{mathtools}
\usepackage{cancel}
\usepackage[table]{xcolor}
\usepackage{colortbl}
\usepackage{algorithm2e}
\usepackage{appendix}

\usepackage{amsthm}
\theoremstyle{definition}
\newtheorem{defn}{Theorem}

\definecolor{lgray}{RGB}{210,210,210}

\begin{document}

\newcommand{\mean}[1]{\left\langle #1 \right\rangle} 
\newcommand{\abs}[1]{\left| #1 \right|}

\title{Intervention scenarios to enhance \\ knowledge transfer in a network of firms}   

\titlealternative{Intervention scenarios to enhance knowledge transfer in a network of firms}   

\author{Frank Schweitzer, Yan Zhang, Giona Casiraghi}

\authoralternative{F. Schweitzer, Y. Zhang, G. Casiraghi}

\address{Chair of Systems Design, ETH Zurich, Weinbergstrasse 58, 8092, Zurich, Switzerland}

\reference{(Submitted for publication: 25 June 2020)}
\www{\url{http://www.sg.ethz.ch}}

\makeframing
\maketitle
\begin{abstract}

  We investigate a multi-agent model of firms in an R\&D network.
  Each firm is characterized by its knowledge stock $x_{i}(t)$, which follows a non-linear dynamics.
  It can grow with the input from other firms, i.e., by knowledge transfer, and decays otherwise. Maintaining interactions is costly. 
  Firms can leave the network if their expected knowledge growth is not realized, which may cause other firms to also leave the network. 
  The paper discusses two bottom-up intervention scenarios to prevent, reduce, or delay cascades of firms leaving.
  The first one is based on the formalism of network controllability, in which driver nodes are identified and subsequently incentivized, by reducing their costs.
  The second one combines node interventions and network interventions.
  It proposes the controlled removal of a single firm and the random replacement of firms leaving.
  This allows to generate small cascades, which prevents the occurrence of large cascades.
  We find that both approaches successfully mitigate cascades and thus improve the resilience of the R\&D network.

\end{abstract}

\section{Introduction}

Interventions belong to the tool box of \emph{systems design} \citep{Schweitzer:2019vp}.
The ability to influence systems such that they reach a desired state or show a desired behavior, is not only of relevance for engineers and operators.
This ability is also favored by managers or politicians, who wish to steer the dynamics of socio-economic systems towards a particular outcome.
Most of the interventions in economic systems are targeted at the macro level, for instance by adjusting tax rates or legal conditions.
They follow a \emph{top-down} approach: a centralized decision to change some ``boundary conditions'' induces an adaptation of the system, hopefully in the right, i.e., wanted, direction.

This approach is contrasted with the \emph{bottom-up} approach that targets system elements rather than systems as a whole~\cite{Roberts2011,Borgatti2006,Heckathorn1997}. It plays a major role in complex systems comprising a large number of elements, denoted as \emph{agents} in the following.
Bottom-up interventions can be targeted either at specific agents or at their interactions or at the network as a whole \citep{valente2012network}.
In socio-economic systems, agent specific interventions include, for example, monetary incentives (e.g., reduced costs, bonuses) or privileged access to resources (e.g., information, credit)~\cite{LeoneSciabolazza2020,Whetsell2020}. 

Complex socio-economic systems are often represented as networks, where agents are depicted as nodes and their interactions as links.
To utilize a bottom-up approach of interventions requires to solve a number of problems that are later addressed also in this paper.
First of all, we need to find out which agents are the most promising ones to drive a system, i.e., we need to identify \emph{driver nodes}. 
This problem can be only tackled if we know how agents influence another, i.e., we need to specify the \emph{dynamics \emph{on} the network}.
Secondly, we need to determine the desired state of the system, and eventually we need to calculate the appropriate intervention that is suitable to drive the network towards this state.

These problems are exacerbated if, in addition to the dynamics \emph{on} the network, there is also a dynamics \emph{of} the network.
That means, the network itself changes because (i) agents join or leave the network or (ii) add, remove, or rewire their links to other agents. 
The dynamics of the system is then described by two time scales, one for the interactions, and one for the change of the network.
Most real-world socio-economic systems are characterized by these couplings of time scales.
For example, online social networks become quite volatile because users enter and exit and also change their interactions with other users at high frequency.
This makes it difficult to assess the robustness of such systems.
If users start to leave and this way generate large cascades of other users leaving, these networks may even collapse \citep{Garcia2013d}. 

To model the direct and indirect impact of nodes leaving a network, two different approaches are followed. 
The first one focuses on the network \emph{topology}, specifically on the degree of the nodes and their embedding.
One example is the $k$-core decomposition model \citep{Baxter2012b,garas2012b} to explain cascades.
The second one also includes the internal dynamics of the nodes, for example the \emph{threshold} model \citep{Hackett2011,Watts2002}.
Combinations of these approaches, such as the $KQ$ model \citep{Yu2016} lead to a better formal understanding of the \emph{emergence} of \emph{systemic risk} \citep{lorenz2009b,burkholz2018,burkholz2016}, i.e., the collapse of a large part of the system because of failure cascades.

Also economic networks are prone to failure cascades and decline \citep{SCHWEITZER2009,saavedra2008asymmetric}. 
In this paper, we discuss the case of \emph{R\&D networks}, used for the knowledge exchange between firms.
Empirical studies have shown that such networks exhibit a \emph{life cycle dynamics}, i.e., they grow and later decay \citep{tomasello2016,powell2005nda} because
in R\&D collaborations  firms usually terminate their interactions after some time.
Either the purpose of their collaborations is fulfilled, e.g., a number of patents are filed, or it is \emph{not} fulfilled.
For example, firms have not obtained an expected knowledge stock or an expected growth of knowledge and therefore decide to leave the network to save the costs involved in collaborations. 
If firms leave, this may cause other firms to leave as well, because they lost the input from their collaboration partners.
Thus, before we can think of interventions, we need to model this process, which is  provided in Section~\ref{sec:model}.

Based on these insights, we will then develop two different approaches towards interventions in Sections~\ref{sec:interv-strat-driver} and \ref{sec:comp-two-interv}.
The aim of these interventions is to prevent large cascades of firms leaving the collaboration network.
If that is not entirely possible, we want at least to reduce the size of such cascades, or to delay them. 
Our first approach is rooted in control theory \citep{Haynes1970,Sussmann1972} applied to complex networks, a recent development that lead to the concept of network controllability \citep{Liu2011d}.
Basically, we identify a set of driver nodes with two different methods, to which an incentive is applied. 
The formal method relies on a \emph{linear} dynamics on the network, which is often not given.
Generalizations to  non-liner problems are not straightforward, as discussed in \cite{Whalen2015,Cornelius2013}.
For these reasons, we will mainly use computational methods in this paper.

To contrast the formal approach, our second approach is based on heuristics, that means on experience and intuition.
It combines two different types of interventions: a \emph{network} intervention, targeted at the whole network to allow a continuous evolution, and a \emph{node} intervention targeted at only one firm. 
It was already shown, for the simpler case of a linear dynamics, that such heuristic approaches can significantly improve the robustness of networked systems \citep{Casiraghi2019a}.
Here we investigate the case of a non-linear dynamics and a fixed driver node.
We are mainly interested to see how this heuristic intervention fares in comparison to the formal intervention based on network controllability.
What are the advantages and shortcomings of these two different interventions?
And to what extent can they be applied to a knowledge exchange network of firms?
These questions will be discussed in Section~\ref{sec:conclusions}.

\section{Generation of Knowledge stock}
\label{sec:model}
\subsection{A network model of interacting firms}
\label{sec:netw-model-inter}

\paragraph{Knowledge stock. \ }

In the following, we utilize a multi-agent model of firms $i=1,...,N$, which are each characterized by a scalar variable $x_{i}(t)$, their \emph{knowledge stock}.
This summarizes for example the R\&D (research and development) experience of a firm, measurable by its number of patents and research alliances.
$x_{i}$ can have continuous values which have to be positive, and can change over time at a time scale $t$. 

To specify this dynamics we first note that the value of the knowledge stock continuously decreases if it is not maintained.
Hence, we consider a decay term $-\gamma x_{i}(t)$ with a decay rate $\gamma$ characterizing the life time of the knowledge stock.   
To compensate for the decay, we need to make assumptions about the growth of the knowledge stock.
One could reasonably argue about a source term that reflects \emph{inhouse R\&D activities} \citep{gassmann1999new}. 
The main focus in our paper, however, is on knowledge \emph{transfer}.
Therefore, we assume that the growth of $x_i$
is mainly driven by input from \emph{other firms}, i.e., by \emph{R\&D collaborations}, rather than by own research activities.
This reflects empirical observations for \emph{innovation networks} of firms \citep{tomasello2016}.

\paragraph{Network representation. \ }

To model knowledge transfer, we use a \emph{network}, in which \emph{nodes} represent agents, i.e., firms, and \emph{links} their interactions.
The network approach implies that interactions in a group of $n$ firms, e.g., in R\&D alliances between 2 to 10 firms, are decomposed into \emph{dyadic interactions} between any two of these $n$ firms.
We further consider \emph{direct interactions} between firms.
I.e., an interaction $i \to j$ describes that firm $i$ transfers knowledge to firm $j$,
but this does not necessarily has to be \emph{reciprocal}, i.e., $j \not \to i$ is possible.
With $N$ firms, there are $N(N-1)$ different directed dyadic interactions possible.
Whether they take place is described in an \emph{adjacency} matrix $\mathbf{A}$ of size $N\times N$, in which an entry $a_{ij}=1$ indicates a link $i\to j$, and $a_{ij}=0$ its absence.
Empirical investigations have shown that such matrices for R\&D collaborations are usually \emph{sparse}, that means the number of links is of the order $N$ rather than $N^{2}$. 
Based on the entries in the adjacency matrix we can define the \emph{in-degree} $d^{+}_{i}=\sum_{j}a_{ji}$ of a firm $i$ as the number of incoming links from other firms, and the \emph{out-degree} $d^{-}_{i}=\sum_{j}a_{ij}$ as the number of outgoing links to other firms. Note that in general $a_{ij}\neq a_{ji}$. 

\paragraph{Knowledge growth. \ }

Our main assumption is that the knowledge growth of firm $i$ is determined by the knowledge stock of those firms $j$ that have \emph{direct link} to $i$, as expressed by the $a_{ji}$, i.e., it is proportional to $x_{j}(t)$, with the proportionality constant $b$ to weight the benefit.
Further, it is reasonable to consider a \emph{saturation} for the growth of knowledge stock. 
At higher levels of $x_{i}$, it becomes more difficult for firm $i$ to ``absorb'' new knowledge, i.e., to incorporate it into a firm because of the internal complexity associated with the way knowledge is stored and linked internally.
This absorptive capacity is for simplicity reflected in a quadratic term $\hat{c}\ x_{i}^{2}$, as known from saturated growth dynamics \citep{schweitzer2020}.

The proportionality factor $\hat{c}$ for the saturation effect, however, deserves a discussion, as there are different arguments possible.
It could be simply a constant, $\hat{c}\equiv c_{0}$, to reflect that the saturation effect depends mainly on the \emph{knowledge stock} of firm $i$.
On the other hand, it could be also an individual variable, $\hat{c}\equiv c_{i}$, to incorporate further influences.
One possibility is a dependency on the \emph{in-degree}, $c_{i}^{+}=c d_{i}^{+}$, based on the argument that a larger \emph{number} of firms transferring knowledge to firm $i$ makes the absorption of knowledge more difficult.
Another possibility is a dependency on the \emph{out-degree}, $c_{i}^{-}=c d^{-}_{i}$, based on the argument that, in addition to the absorption of knowledge, firm $i$ also has to maintain links to other firms $j$, which consumes resources, and therefore exacerbates the problem.
In this paper, we follow the latter argumentation which was also used in \citep{Koenig2009a}, i.e., $\hat{c}=cd^{-}_{i}$.  

Combining these assumptions, we propose the following dynamics for the knowledge stock \citep{Koenig2009a}:
\begin{equation}
  \frac{dx_{i}(t)}{dt}= - \gamma\, x_{i}(t) + b \sum\nolimits_{j} a_{ji}\, x_j(t) + b^{\mathrm{ext}} \sum\nolimits_{j} p_{ji}\, x_j(t)
  - c d_{i}^{-} x_{i}^{2}(t) \label{eq:18}
\end{equation}
Here, it is additionally considered that some links, denoted by $p_{ji}$, provide firm $i$ with a direct input from particular valuable firms.
For example, instead of obtaining \emph{indirect} knowledge input from a firm $k$ via other firms $j$, firm $i$ would much more benefit if $k$ had a \emph{direct} link to $i$.
So, if $p_{ji}=1$, there will be an extra benefit  $b^{\mathrm{ext}}$ from interacting with this valuable agent.
Such shortcut externalities have been discussed e.g., in \citep[p.247]{Koenig2009a}.
In the following, we drop this term in the dynamics, to simplify the analysis.

\begin{figure}[htbp]

\begin{subfigure}{.3\textwidth}
    \includegraphics[height=3cm]{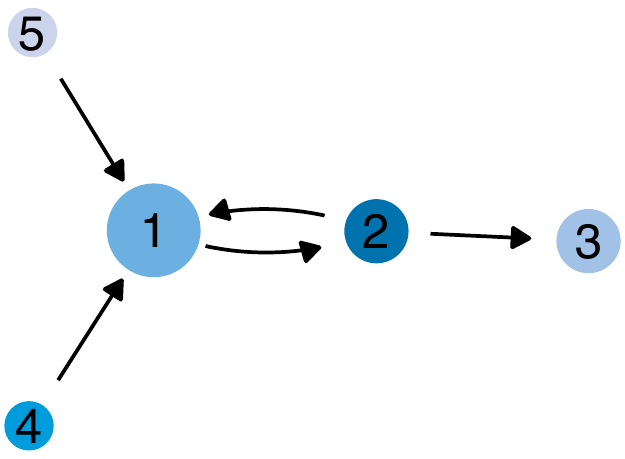}

    \centerline{\includegraphics[width=.99\textwidth,angle=0]{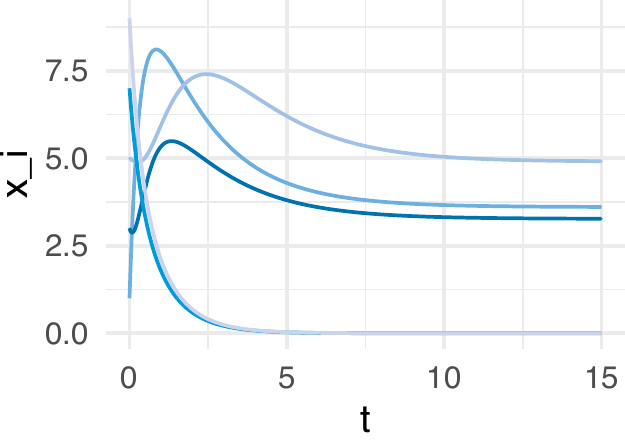}} \caption{}
  \end{subfigure}\hfill\vline\hfill
  \begin{subfigure}{.3\textwidth}
    \includegraphics[height=3cm]{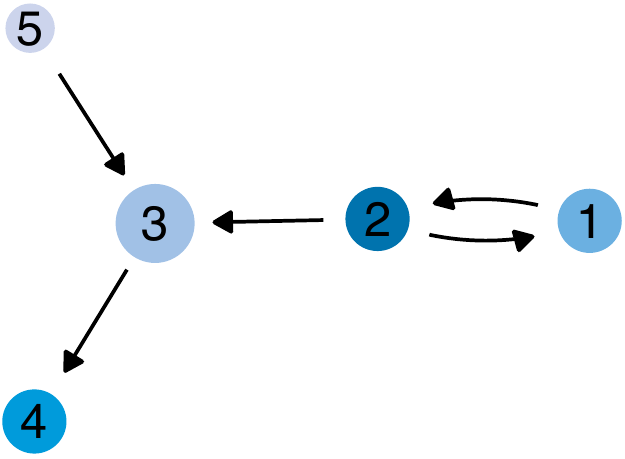}
    
    \centerline{\includegraphics[width=0.99\textwidth,angle=0]
        {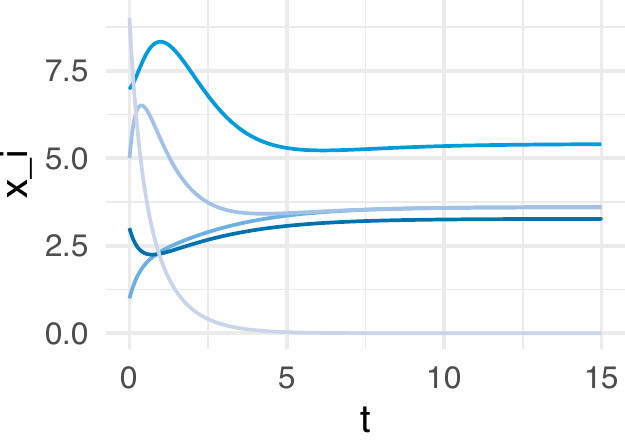}} \caption{}
\end{subfigure}\hfill\vline\hfill
\begin{subfigure}{.3\textwidth}
 \includegraphics[height=3cm]{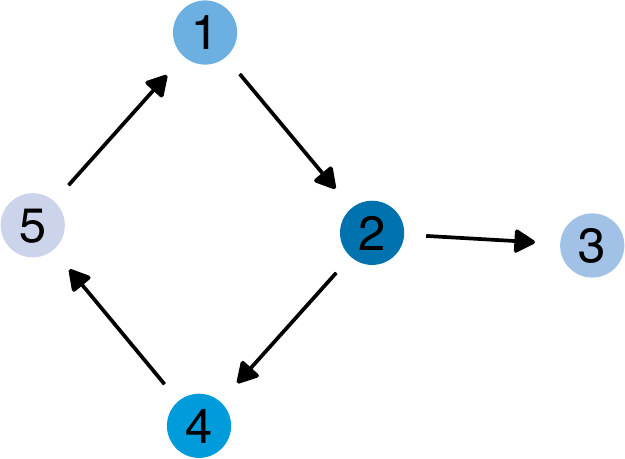}

    \centerline{\includegraphics[width=0.99\textwidth,angle=0]
        {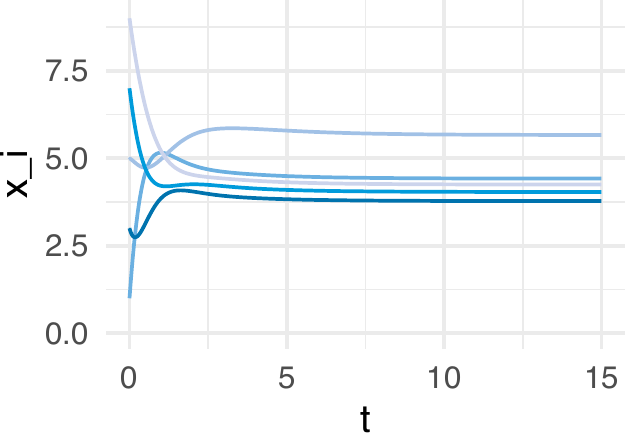}} \caption{}
\end{subfigure}

  \caption[Reputation Dynamics for 3 4-nodes graphs]{Dynamics of the knowledge stock, Eqn.~\eqref{eq:18}, in a small network of 5 firms.
The size of a node is proportional to its in-degree. 
}
  \label{fig:simplereputation}
\end{figure}

\paragraph{Direct and indirect reciprocity. \ }

The above dynamics leads to a stationary, but non zero value of the knowledge stock only if the network contains \emph{cycles} of firms benefiting another, as illustrated in Figure~\ref{fig:simplereputation}. 
The shortest possible cycle involves two firms, for example $1\to 2$, $2\to 1$ in Figure~\ref{fig:simplereputation}(a,b).  
This mutual interaction relates to \emph{direct reciprocity}, as firm 1 contributes to the knowledge stock of firm 2 and the other way round.
But these cycles can be also larger, as  Figure~\ref{fig:simplereputation}(c) illustrates.
In this case, firm 1 still contributes to the knowledge stock of firm 2, but does \emph{not} receive a reciprocal benefit from firm 2.
Still, firm 1 benefits from being part of a closed cycle $1\to 2 \to 4 \to 5 \to 1$, i.e., it receives an indirect benefit from firm 2, via firms 4 and 5.
Such constellations relate to  \emph{indirect reciprocity}, which plays an important role also in the emergence of cooperation \citep{nowak1998dynamics}.
Also the development of technology depends on the existence of such  feedback cycles \citep{abrahamson1997social}. 
Hence, links that contribute to \emph{closing} these \emph{cycles} in the interaction network could for example receive an extra benefit, $b^{\mathrm{ext}}$, although this is not considered here.

It is worth noticing that not \emph{all} firms need to be part of cycles, to obtain a non-zero knowledge stock.
In Figure~\ref{fig:simplereputation}(a), we show a configuration where firm 3 benefits from the knowledge transfer from firm 2, but does not contribute in a reciprocal manner.
Because this even saves the costs from maintaining the links, firm 3 obtains the highest value $x_{3}$.
Comparing  Figure~\ref{fig:simplereputation}(a) with Figure~\ref{fig:simplereputation}(b), we see that firm 3, even with a higher in-degree, does not necessarily have the highest value $x_{3}$.
This points to the impact of the \emph{non-linear dynamics} of the knowledge stock, which cannot be simply reduced to the influence of in-degrees.  

Eventually, we note that firms reach a non-trivial stationary knowledge stock only if they are connected to a cycle. 
In Figure~\ref{fig:simplereputation}(a) Firms 4 and 5 contribute to the knowledge stock of firm 1, but do not receive any reciprocal contribution.
Hence, $x_{4}\to 0$ and $x_{5}\to 0$ over time. 

In general, the existence of non-trivial steady states, i.e., $x_{i}(t\to\infty)>0$, for firms $i$ that are part of a network $\mathcal{G}$,
$i\in V(\mathcal G)$, can be ascertained by studying the irreducibility of the graph $\mathcal G$ \cite{jain1998autocatalytic, Koenig2009a}.
In particular, the existence of at least one
closed cycle in the graph is sufficient to assure that both firms in the cycle and firms that receive links from firms within the cycle
have a non-zero stationary knowledge stock. 
Moreover, for special classes of graphs it is possible to find
analytical solutions for the values of the steady states \citep{Koenig2009a}.

\subsection{Decision to leave the network}
\label{sec:rewiring-links}

\paragraph{Knowledge stock and growth. \ }

Firms participate in knowledge transfer because they have the expectation that their knowledge stock grows over time.
So, it is reasonable to assume that after the knowledge stock dynamics has converged to the stationary value, they evaluate the performance.
Based on this, they may decide to \emph{leave} the network, i.e., to cut all their outgoing links.
Because this decision is based on the stationary values of the knowledge stocks, $x_{i}^{\mathrm{stat}}$, it takes place at a different time scale $T$, which is longer than the time scale $t$ at which the knowledge stock reaches its stationary value. 

Specifically, we consider that at each time step $T$, firms compare (A) their absolute value of the knowledge stock, $x_{i}^{\mathrm{stat}}(T)$ with a threshold, $x^{\mathrm{thr}}$ (equal to all firms), and (B) their \emph{growth} of the knowledge stock in the past time step, $g_{i}(T)=x_{i}^{\mathrm{stat}}(T)-x_{i}^{\mathrm{stat}}(T-1)$ with a threshold, $g^{\mathrm{thr}}$ (equal to all firms). 
They decide to continue to collaborate only if both of their respective values are above the thresholds. 
If any of these two conditions are not met, firms decide to leave with a certain \emph{probability} $p$, which implies that they do \emph{not immediately} leave.

We note that both conditions reflect different aspects that influence firms decisions. 
Condition (A) uses a cost/benefit assessment.
If, in the current situation, the costs are higher than the benefits from receiving knowledge input, firms may choose to leave.
That means, higher values of $x^{\mathrm{thr}}$ consider a sensitivity for higher costs. 
Condition (B), on the other hand, captures to what extent the \emph{expectations} of firms regarding the growth of knowledge stock are met.
Higher values of $g^{\mathrm{thr}}$ therefore indicate a lower tolerance of firms, if their expectations are not met.

\begin{figure}[htbp]
\includegraphics[width=0.30\textwidth]{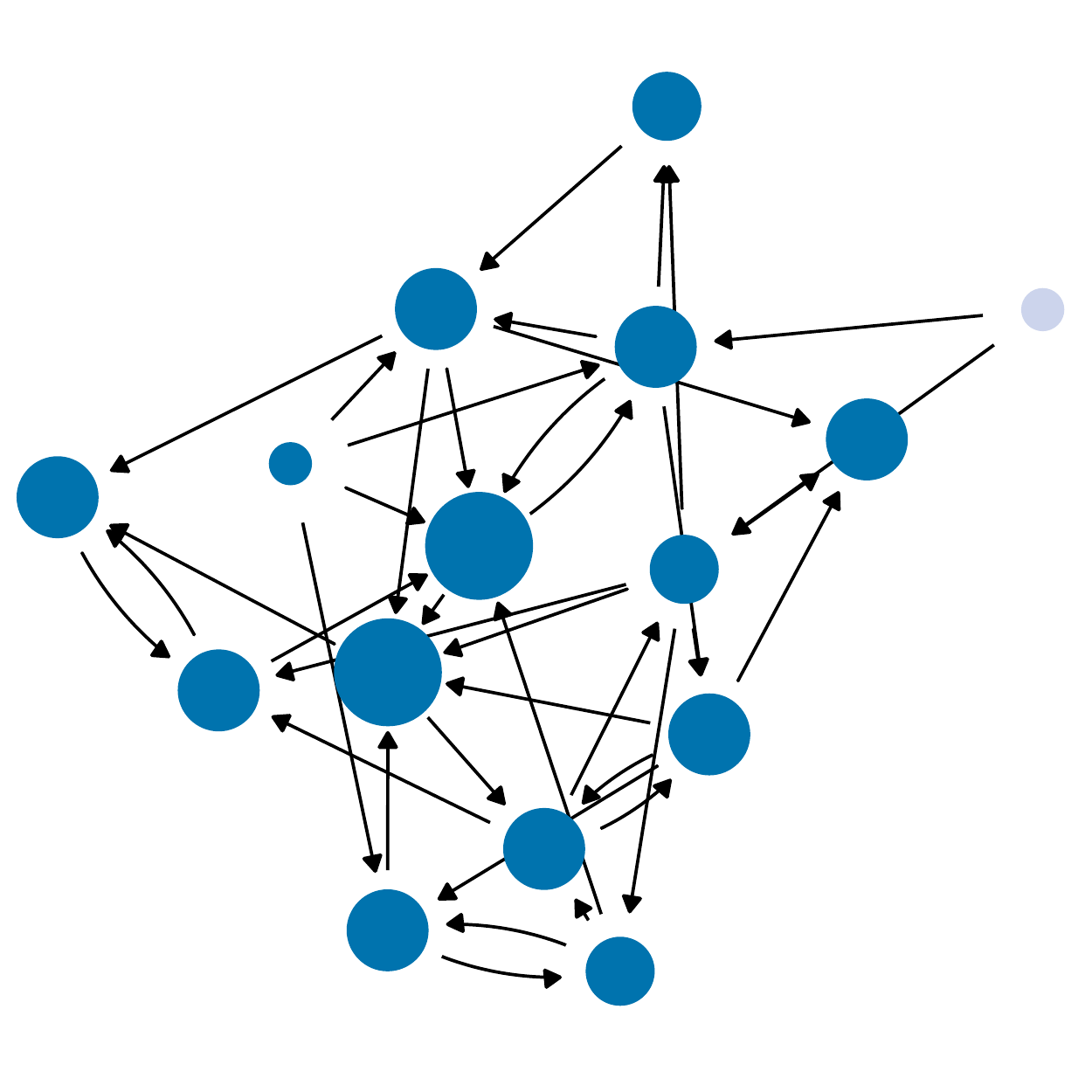} \hspace*{-1cm}$T=1$
\hfill 
\includegraphics[width=0.30\textwidth]{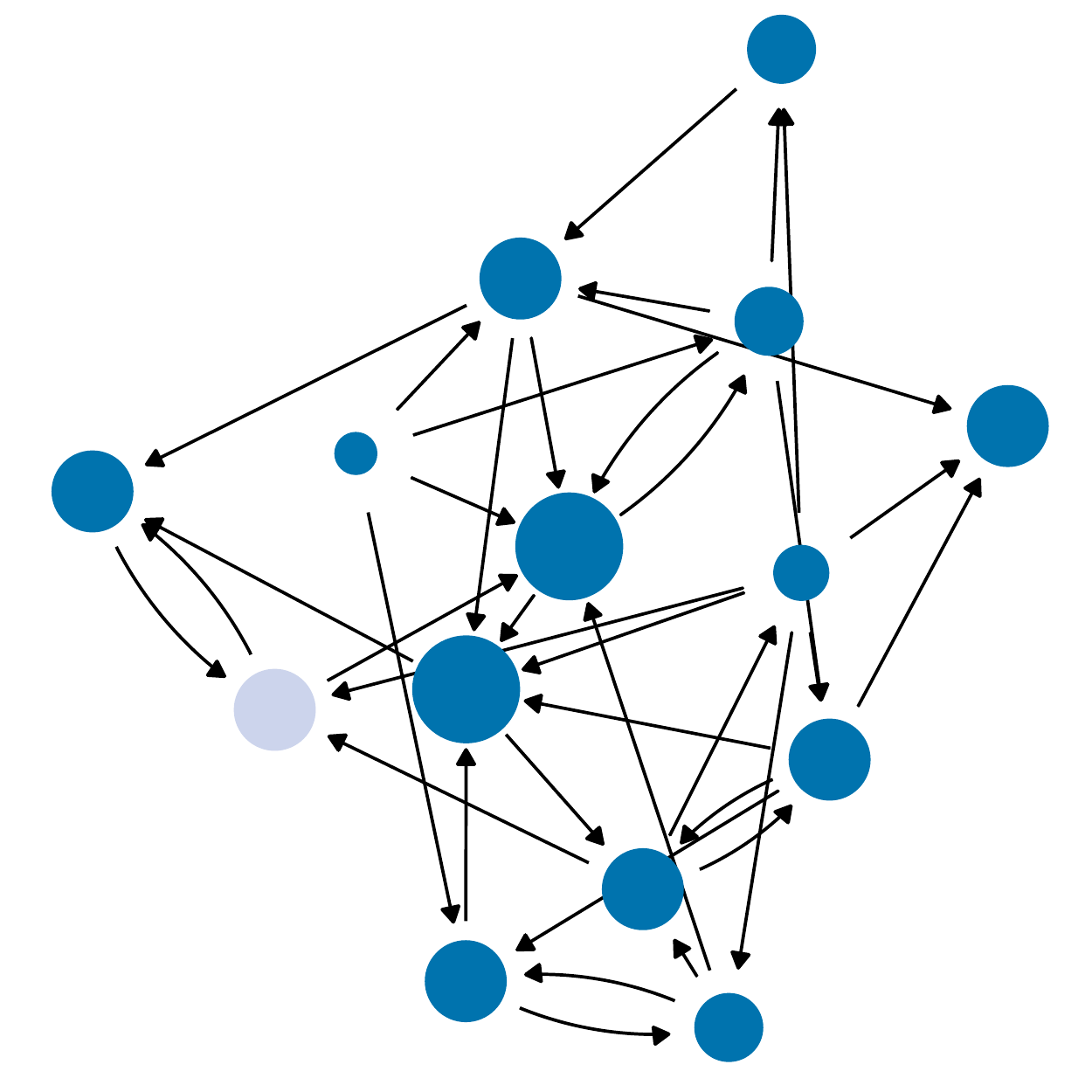} \hspace*{-1cm}$T=2$
\hfill
\includegraphics[width=0.30\textwidth]{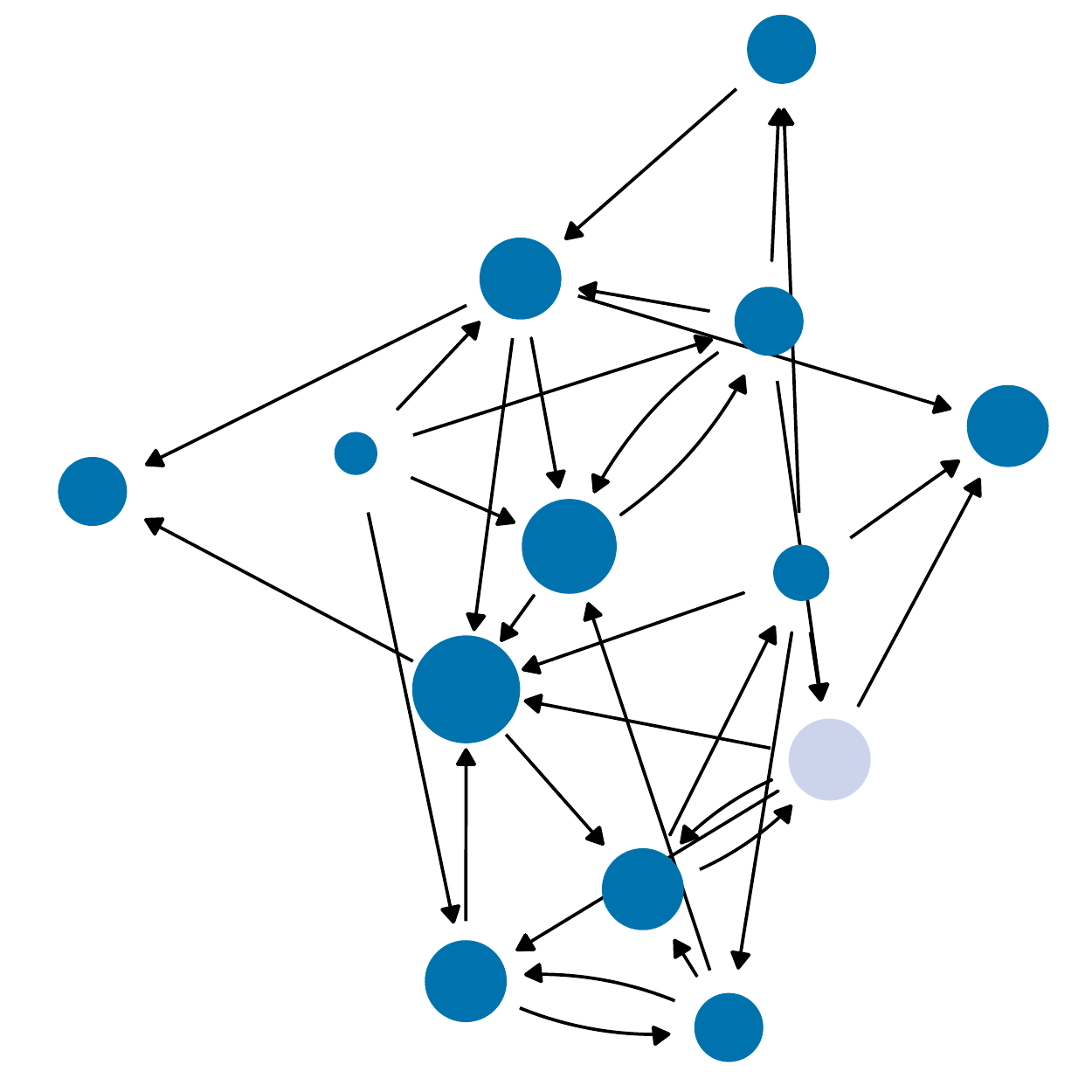} \hspace*{-1cm}$T=3$

\vspace*{1cm}

\includegraphics[width=0.30\textwidth]{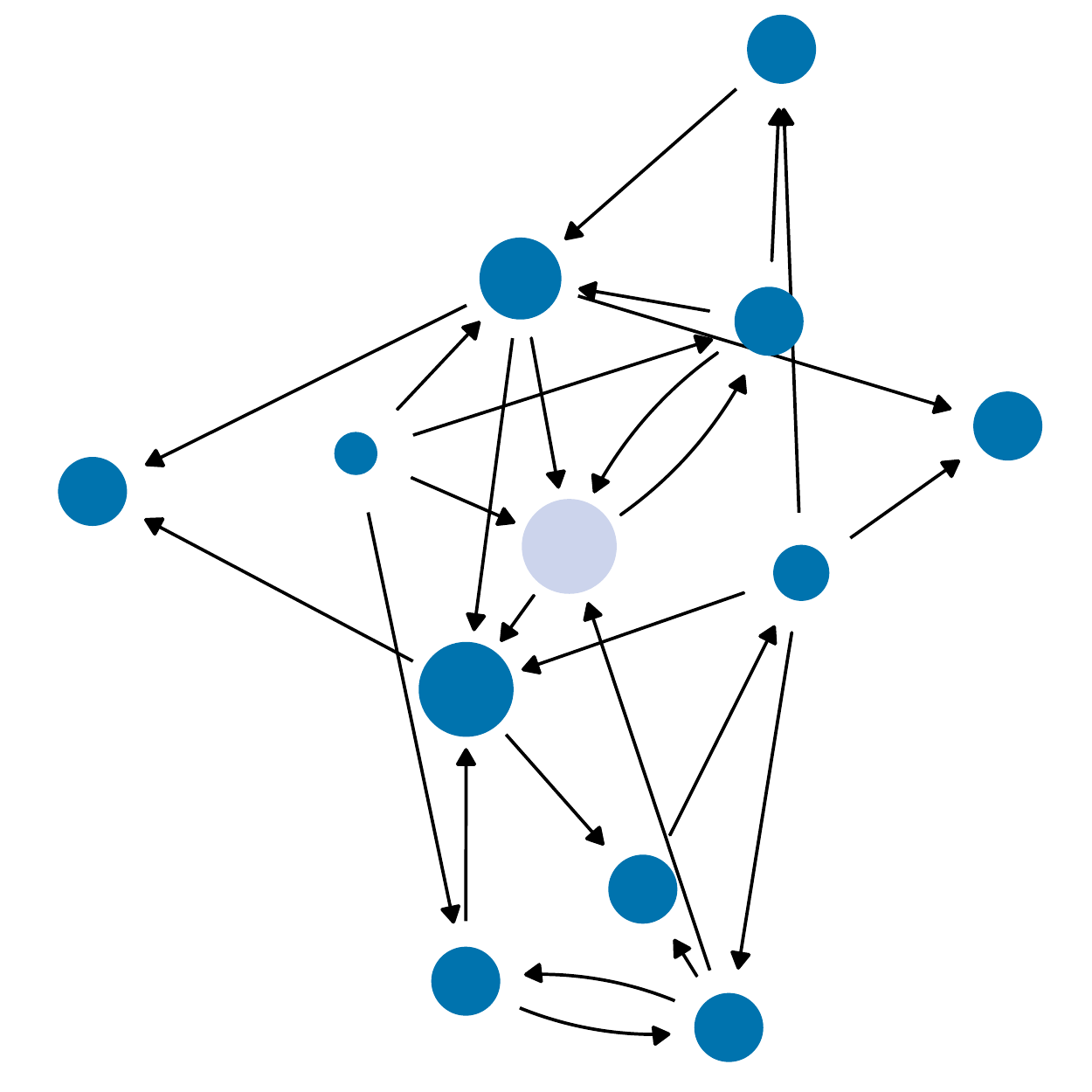} \hspace*{-1cm}$T=4$
\hfill
\includegraphics[width=0.30\textwidth]{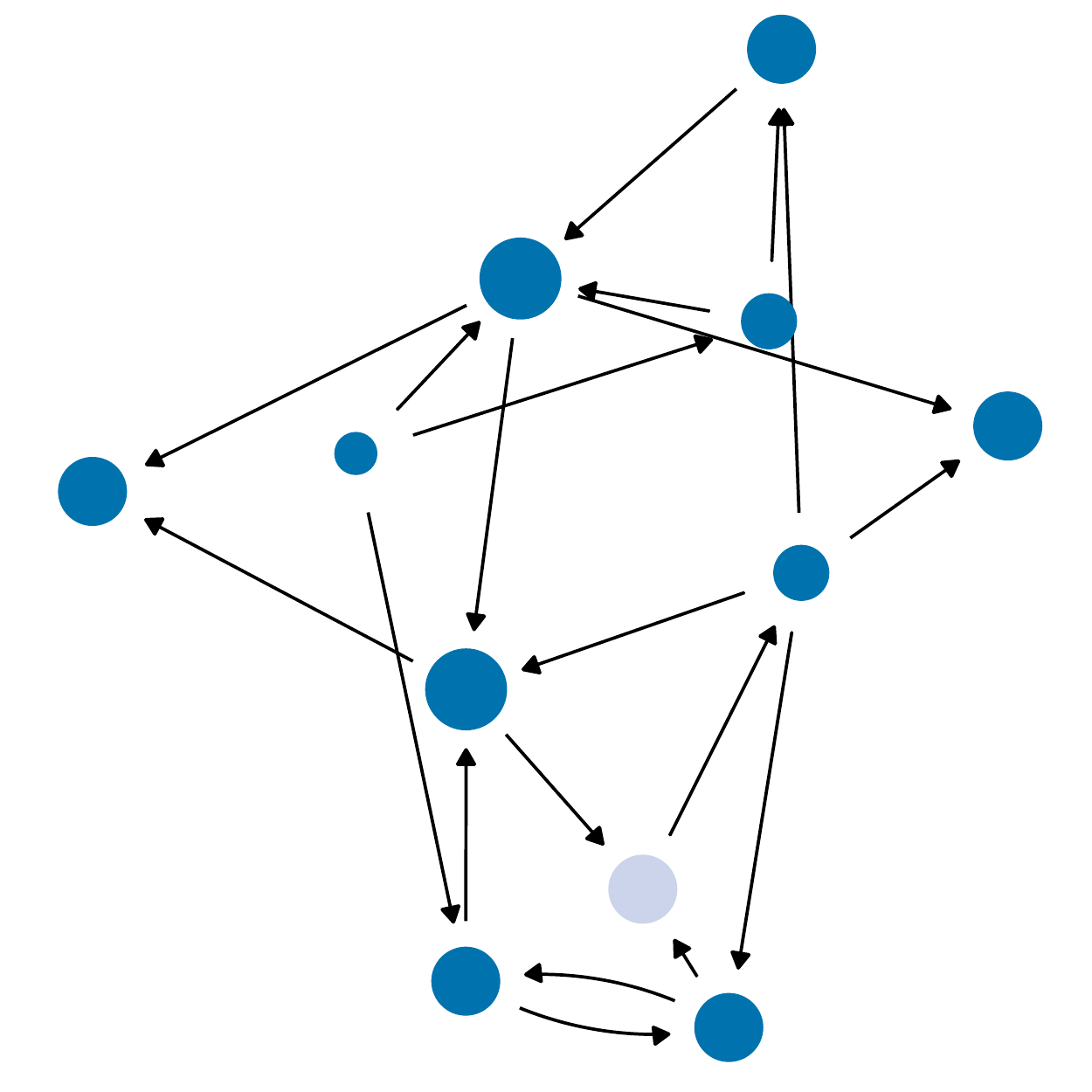} \hspace*{-1cm}$T=5$
\hfill
\includegraphics[width=0.30\textwidth]{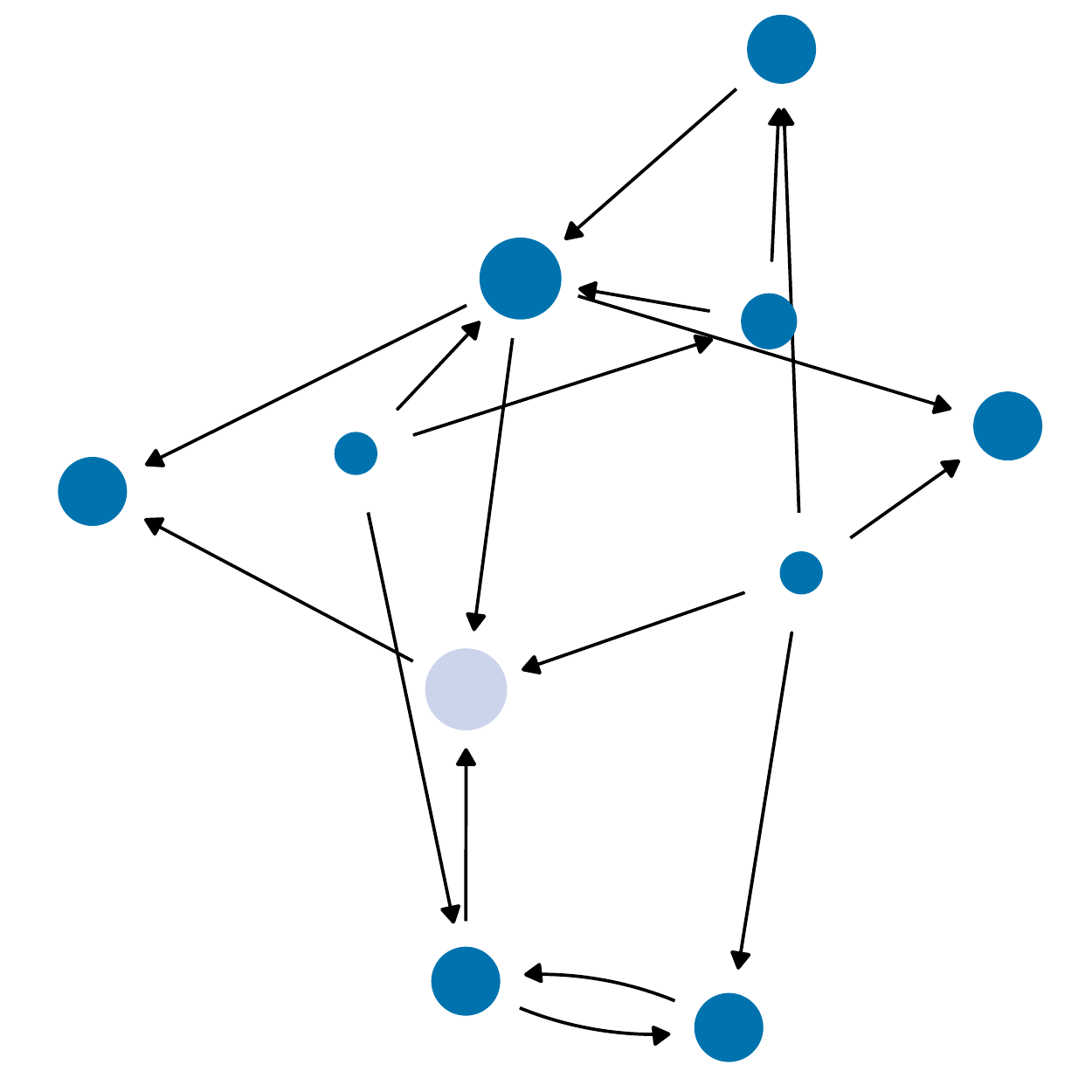} \hspace*{-1cm}$T=6$

        \caption{Cascades of firms leaving a sample network of $N=15$ without any network intervention. The position of the nodes is fixed and their size proportional to their current in-degree.
          Firms colored in gray will leave the network in the next time step $T$. Parameters: $\gamma=0.5$, $b=0.2$, $c=0.06$,  $x^{\mathrm{thr}}=0.05$, $g^{\mathrm{thr}}=0.8$, $p=0.1$.
          }
  \label{abm:figure1}
\end{figure} 

If firms leave the network, this impacts the transfer of knowledge between firms in two ways: (i) directly, because firms leaving no longer contribute to the knowledge stock of their previous partners, (ii) indirectly, because firms leaving change the structure of the network and this way also the cycles of indirect reciprocity.
This, on the other hand, generates an impact on the knowledge growth and the knowledge transfer of the remaining firms in the \emph{next} time step.
Hence, at time $T+1$ the remaining firms obtain a different knowledge stock, $x^{\mathrm{stat}}(T+1)$ and a different growth rate, $g(T+1)$.
This may cause other firms to leave the network, i.e., we observe \emph{cascades} of firms dropping out.
Such observations are in qualitative agreement with empirical studies about the \emph{life cycle} of R\&D network, where indeed a decline of firms participating can be found \citep{tomasello2016}. 

\paragraph{An example. \ }

Figure~\ref{abm:figure1} illustrates such a cascade of firms leaving the network.
At time steps $T=1,5$, the firms colored in gray leave the network because their knowledge stocks $x_{i}^{\mathrm{stat}}(T)$ fell below the threshold value $x^{\mathrm{thr}}$ (condition A).
At time steps $T=2,3,4,6$, on the other hand, the firms colored in gray leave the network because their expectations about their knowledge growth $g_{i}^{\mathrm{stat}}(T)$ fell below the corresponding threshold $g^{\mathrm{thr}}$ (condition B).
These firms have indeed a high knowledge stock, i.e., they pass condition (A).
That means, with just the comparison of knowledge \emph{stock}, we would \emph{not} observe that a cascade emerges.
The comparison of knowledge \emph{growth} thus adds a new element to the dynamics.
Because \emph{expectations} are considered, we observe in our model that also firms with a high knowledge stock and a high number of connections from partners may decide to leave the network, which makes our model different from other cascade models \citep{Watts2002,Hackett2011,burkholz2016}.

\section{A formal approach to interventions}
\label{sec:interv-strat-driver}

\subsection{Interventions and network topology}
\label{sec:nodelevel}

The above discussion makes very clear that  our interventions shall prevent the \emph{breakdown} of the network of firms collaborating in knowledge exchange.
As our ultimate aim,  we want to (i)
\emph{prevent} cascades of firms leaving the network or, if that is not entirely possible, (ii) \emph{reduce} the cascade size such that the majority of firms still remains in the network, or (iii) \emph{delay} the occurrence of cascades such that the network remains at least for a given time horizon, $\mathcal{T}$.

This closely relates to investigations of \emph{systemic risk}, which is measured by the fraction $X(t)$ of failing agents/nodes in a system.
Systemic risk usually \emph{emerges} from the early failure of very few agents, which affects other agents and causes them to fail, this way leading to a \emph{failure cascade}.
If $X(t)$ approaches 1, most parts of the system are destroyed.
Therefore, we can quantify the aim of our intervention strategies as keeping $X(t\to \mathcal{T})$ low, precisely below 0.5.
This means, the \emph{majority} of firms \emph{stays} in the network for a finite, but large time $\mathcal{T}$.

This aim can be achieved by interventions on the \emph{network level} and/or the \emph{node level}.
In the following, we discuss two scenarios that combine interventions on these levels in different ways. 
We start with addressing the \emph{node} level.
Here, the decisions of individual firms to leave is influenced via their knowledge stock, Eqn. \eqref{eq:18}.
A trivial approach would be to simply lower the costs $c$ of \emph{all} firms, this way increasing their knowledge stock and forcing them to stay.
Another trivial variant would be to reduce the probability $p$ of firms to leave if their conditions (A) or (B) to stay in the network, are no longer met. 
We do not follow these trivial approaches which treat all firms equal.
Instead, we want to make use of an important feature that distinguishes firms, their topological \emph{position} in the collaboration network. This shall help us to identify \emph{which firms} should be subject to an intervention.

Large-scale empirical studies of R\&D collaboration networks \citep{tomasello2016} have shown that their \emph{topology} is characterized by two  distinct features: (i) a very broad \emph{degree distribution} and (ii) a distinct \emph{core-periphery structure}.
Firms belonging to the core are often \emph{hubs}, i.e., they have more and denser connections among each other, while firms in the periphery are only loosely connected to the network (or even disconnected, which is not considered here). 
Thus, node interventions could be targeted at core firms, or at firms from the periphery.
Examples for these interventions have been already discussed in the literature \citep{Zhang2016,Casiraghi2019a}, and their performance depends on many details both regarding the interaction dynamics and the network topology.

In our simulation studies, we will use sample networks exhibiting the mentioned topological properties. 
But instead of testing all possible interventions, in this paper we only choose two particular examples that combine information about the topological position of firms and their knowledge stock, as outlined in the following sections. 

\subsection{Identifying driver nodes}
\label{sec:ident-driv-nodes}

To decide which firms should be targeted by an intervention, we choose the formalism of \emph{network controllability} \citep{Liu2011d,Whalen2015,Cornelius2013}, a research field at the intersection of complex networks and control theory.
It allows to identify a set of \emph{driver nodes}, firms in our case, which are then targeted by an intervention.
Control means that applying a control signal to the driver nodes will allow us to drive the dynamics on the network to a preferred outcome, in our case to values of knowledge stock that would prevent firms from leaving the network.

The formalism of \emph{network controllability} requires to know the dynamics \emph{on} the network, i.e., changes in the state variables of the nodes.
This is, in our case, given by the non-linear dynamics of the knowledge stock of each firm, Eqn. \eqref{eq:18}. 
As long as this dynamics can be linearized, the concept of structural controllability still applies.
If a firm represents a driver node, additionally a \emph{constant control signal} $u_{i}$ is added to the dynamics of Eqn. \eqref{eq:18}.  
This leaves us with the tasks to identify the set of driver nodes and to determine the control signals $u_{i}$ such that the dynamics of the whole network is driven toward the desired state, which shows either (i) no, (ii) small, or (iii) delayed cascades.

To determine the driver nodes, we choose two different approaches:
(a) we identify all firms that belong to the 20\% with the highest knowledge stock $x_{i}^{\mathrm{stat}}$, and
(b) we instead identify all firms that belong to the 20\% with the highest control contribution $\mathcal{C}_{i}$, a measure for the ability of firms to influence others, as described below.
These two different sets of driver nodes are calculated on the \emph{initial network}, i.e., before any firm left. 

To apply ranking scheme (a), we have to calculate the $x_{i}^{\mathrm{stat}}(0)$ for all firms, which have non-zero values only if the network contains cycles (see Sect.~\ref{sec:netw-model-inter}).
For a linear dynamics we could simply obtain the stationary values from solving the \emph{eigenvalue} problem that makes use of the known \emph{adjacency matrix} $\mathbf{A}$ of the network.
Unfortunately, the nonlinearities involved in Eqn. \eqref{eq:18} do \emph{not} allow this procedure.
Instead, for all time steps $T$, including $T=0$, we have to run the dynamics of  Eqn. \eqref{eq:18} at the time scale $t$ and wait until it converges.
Specifically, we have to resort on numerical procedures, such as the standard lsoda FORTRAN ode solver by Linda R. Petzold and Alan C. Hindmarsh, which  is interfaced in most programming languages.
We used the R interface provided by the DeSolve R library.
The computational effort involved was the main reason why we restrict ourselves to the simulation of relatively small networks.
Our main goal is to study the impact of network and node interventions, which can be sufficiently demonstrated with small networks.

Our ranking scheme (b) uses the values of \emph{control contribution} $\mathcal{C}_{i}$ of each firm.
Here we only sketch the idea behind this measure, the details are given in ~\citep{YZ-ACS}.
We first need to identify a minimum set of driver nodes (MDS) required to control the \emph{whole} network. 
Noteworthy, a network of size $N$ can be controlled by different MDS that not always contain the same set of nodes.
Thus, the probability that a node $i$ becomes part of an MDS is denoted as $P(D_{i})$, sometimes called \emph{control capacity}  $\mathcal{K}_{i}$ \citep{Jia2013f}.
Each driver node $i$ in a MDS controls a non-overlapping part of $N_{i}$ nodes of the whole network.
$P(N_{i})$ then denotes the probability that a given node is part of the subnetwork controlled by node $i$.
But we are interested in the conditional probability $P(N_{i}|D_{i})$ that a given node is part of the subnetwork controlled by $i$ given that $i$ is a driver node.
This conditional probability can only be obtained algorithmically as discussed in the mentioned approach of structural controllability.

The upper bound of $P(N_{i}|D_{i})$, normalized by the system size, is called \emph{control range}, $\mathcal{R}_{i}$ \citep{Wang2012e}.
I.e., it gives the maximum relative size of the network controlled by $i$.
Control contribution $\mathcal{C}_{i}$ now \emph{combines} these two different information: $C_{i}=\mathcal{K}_{i}\mathcal{R}_{i}$. 
That means,  $\mathcal{C}_{i}$ of node $i$ captures the probability
for any node in a network to be controlled by node $i$ joint with the probability that $i$ becomes
a driver.
A large value of $\mathcal{C}_{i}$ indicates a large impact of the respective node on driving the whole network to a desired state. 
For an illustrative calculation of control contribution $\mathcal{C}_{i}$ we refer to Ref.~\citep{YZ-ACS}.

It is worth noticing that control contribution $\mathcal{C}_{i}$ is not simply correlated to other topological measures, such as node degree, and thus indeed provides new information to characterize driver nodes.
Further, it was demonstrated on empirical and synthetic networks that identifying top driver nodes by their control contribution rather than by alternative measures such as control range or control capacity, leads to much improved results for network controllability \citep{YZ-ACS}.

If we compare the sets of driver nodes from the two different ranking schemes, we note only a minor overlap in the chosen firms.
This is understandable because a higher knowledge stock is positively correlated with the in-degree of firms, whereas driver nodes chosen by their control contribution mostly have a low in-degree.

\subsection{Results of computer simulations}
\label{sec:results-many}

We illustrate the performance of our intervention scenarios by means of agent-based computer simulations.
These are motivated by the fact that our model involves two time scales: the network structure changes at the time scale $T$,
while the knowledge stock $x_{i}(t)$ of each firm changes at the smaller time scale $t$.
To calculate from Eqn. \eqref{eq:18} the stationary values $x_{i}^{\mathrm{stat}}(T)$, we resort to numerical integration of the set of ordinary differential equations, as mentioned above.  

In our first intervention scenario, we consider a larger number $N_{d}$ of driver nodes, hence, we also simulate a larger network, $N=200$.
The set of driver nodes contains the top 20\% of firms in both ranking schemes discussed in Section~\ref{sec:ident-driv-nodes}, i.e., $N_{d}=40$. 
We recall that the sets of drivers are determined based on the initial network and then kept as drivers.

Because firms decide to leave the network if the conditions (A) or (B) are met, the number of firms in the network can only \emph{decrease}.
The systemic variable $1-X(T)$ then gives us the fraction of firms that remain in the network.
It is of interest to us whether interventions on the driver nodes are able to (i) prevent, (ii) reduce, or (iii) delay a network breakdown.
Our reference case is a scenario without any node interventions.

Whether or nor cascades occur depends, in addition to the network topology, also on the parameters of the model.
We recall that in particular the two thresholds $x^{\mathrm{thr}}$ and $q^{\mathrm{thr}}$ determine the conditions (A) and (B) under which firms decide to leave the network. 
Further, their knowledge stock and knowledge growth depends on the parameters $\gamma$, $b$ and $c$.
All these parameters also impact the control signal $u_{i}$ that is needed for the driver nodes to prevent firms from leaving the network. 

For our first intervention scenario, we have chosen $u_{i}(T)\equiv u(T)=0.1c\;\Theta[X(T)-0.05]$.
That means the intervention is the same for all driver nodes and it is rather small, only 10\% of the cost factor $c$.
Precisely, firms acting as driver nodes still bear 90\% of the costs applied to all firms. 
This is far from a scenario where firms are paid for staying in the network. 
Further, this reduction of the costs is \emph{not} applied continuously to all driver nodes at all times.
The Heaviside function $\Theta[x]$ ensures that the intervention takes place only if the cascade of firms leaving exceeds 5\%, i.e., $X(T)\geq 0.05$.
This takes into account that, in real-world scenarios, there may be a delay in implementing interventions and that it may take time for managers to realize the risk of a potential network crash.

The threshold of 5\% is not independent of the probability $p$ that a firm indeed leaves the network if either conditions (A) or (B) are met.
Smaller values of $p$ result in smaller cascades at a given time $T$, and therefore slow down the process.
We emphasize that this way we do \emph{not prevent} cascades.
If the  conditions (A) or (B) are not met, a firm will not stay in the network, but it may not leave immediately.
Thus the time scale of the cascade is impacted such that, hopefully, the node intervention occurs in time.  

Our choice of parameters shall prevent us from simulating \emph{trivial scenarios}.
Obviously, we can always prevent cascades with (a) a very high level of $u$ and (b) a continuous intervention at driver nodes.
This would immediately finish the paper.
But we are more interested to learn if modest interventions, i.e., small control signals placed at critical times and not continuously, would be able to prevent a network breakdown.
That is why the parameters are chosen such that without interventions a considerable fraction of firms would leave.

 \begin{figure}[htbp]
  \centering
\includegraphics[width=.45\textwidth]{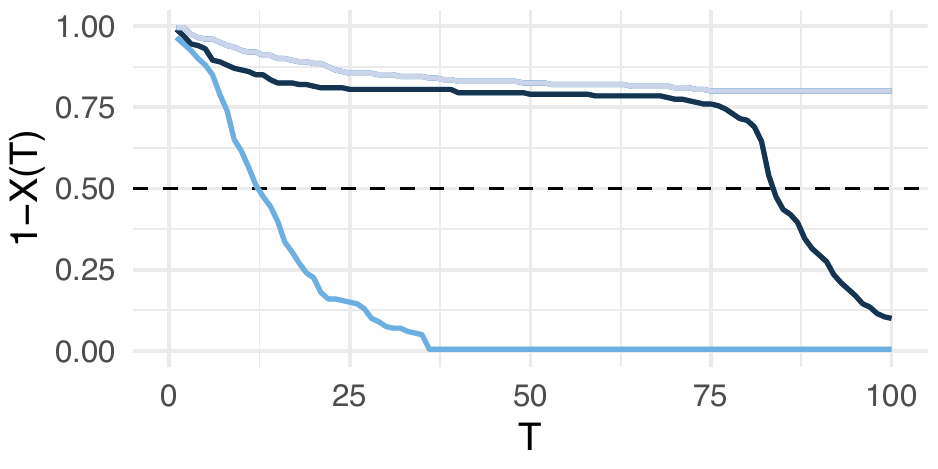}
  \caption{Fraction of firms remaining in the network, $1-X(T)$ over time $T$ for different inter\-ven\-tions.
    Driver nodes are chosen from firms with the highest initial knowledge stock (black) or with the highest initial control contribution (grey). Reference case: no node interventions  (blue). 
    Parameters:  $\gamma$=0.5, $b$=0.2, ${c}$=0.06, $p$=0.1.}
  \label{abm:figure2}
\end{figure} 

Typical results of our node interventions are shown in Figure~\ref{abm:figure2}.
Here we use a fixed time horizon, i.e., $T^{\mathrm{max}}=\mathcal{T}=100$, i.e., we are only interested in the system dynamics during that given period.
This time horizon would, for example, be sufficient for managers to respond to the fact that collaborating firms are leaving. 
We clearly see that, without any intervention, firms are continuously leaving the network which ceases to exist after a short time.
Using driver nodes selected according to their high knowledge stock (ranking scheme a) can considerably improve the situation because large cascades are delayed.
They are, however, not prevented. 
But if driver nodes are chosen according to their control contribution, cascades can be also prevented, at least for the chosen time horizon $\mathcal{T}$=100. 

The interactions of firms and the emerging cascades follow a deterministic dynamics.
But there are random influences in the way the initial network is generated.
For a network of size $N$, firms randomly connect to $N-1$ other firms with a certain probability $q$.
The topology obtained this way also impacts the cascade dynamics.
In order to account for this, we have run 100 simulations to find out, when the majority of firms have left, i.e., from each run we determine $\hat{T}$ from the condition $1-X(\hat{T})\leq 0.5$ and then average over $\hat{T}$.

Our results show that, without any intervention,
the average value is ${\hat{T}^{\mathrm{av}}_{\mathrm{no}}}=36.7$, but the median is $\hat{T}^{\mathrm{md}}_{\mathrm{no}}$=17. 
That means in half of all cases $\hat{T}$ was smaller or equal to only 17, which points to a rather left skew distribution and a fast breakdown. 

\emph{With} our interventions, we are able to considerably mitigate this dynamics.
Using \emph{control contribution} as the best ranking scheme, consistently, the average value increases by $\Delta \hat{T}^{\mathrm{av}}_{\mathrm{cc}}=40.1$ compared to the case of no interventions, i.e., it doubles. 
More interesting, the median strongly increases, and in more than half of the runs the maximum  time horizon $\mathcal{T}=100$ is reached.
We used a one-sided Wilcoxon rank sum test with continuity correction for 100 observations, to show that this improvement is statistically  significant (p-value $\approx5.6e-15$).

Using knowledge stock as the ranking scheme for driver nodes also delays large cascades significantly
compared to the case without interventions (p-value $\approx1.86e-12$).
But both the average and the median values are slightly smaller than with control contribution as ranking scheme.
At the same time, we note that the differences between the two ranking schemes is not really significant (p-value $= 0.092$).
Taking into account the considerable effort in computing the initial control contribution $\mathcal{C}_{i}(0)$ for all firms, we could argue that this effort does not pay off, at least not for the given set of parameters. 

Thus, the main difference in mitigating cascades is between \emph{no} intervention and intervention.
For both ranking schemes, the total \emph{size} of the cascades becomes significantly \emph{smaller}, i.e., more firms remain in the network, and the occurrence of cascades is also \emph{delayed}.
That means, while we cannot completely prevent cascades of firms leaving, given our set of parameters, we can still reach the aim of our interventions, namely to  reduce and to delay them.

\section{A heuristic approach to interventions}
\label{sec:comp-two-interv}

\subsection{Combining node level and network level interventions}
\label{sec:rand-interv-strat}

While the formal approach to node interventions was quite successful, as illustrated above, it also has a number of drawbacks.
Cascades were only reduced and delayed, therefore the number of firms in the network necessarily decayed over time.
Further, the intervention had to target a larger number of firms.
This implies a considerable effort because interventions are costly.
First of all, in a real economic network we need to get access to the firms we want to use as driver nodes, and secondly, 
the control signal itself is also costly.
We remind that $u$ is used to lower the costs of those firms that act as driver nodes.

Thus, we would like to have an intervention scenario that uses only a small number of drivers, ideally only one firm, and that compensates for the loss of firms from cascades that could not be prevented.
Such a scenario is laid out in the following, and it combines interventions both on the node and on the network level. 
We call this intervention \emph{heuristic} because it is not derived in a  formal manner but based on intuition and experience.
As we will show, it works surprisingly well.

We start by describing the  intervention on the \emph{network} level, or systemic level, which shall be used to compensate for the loss of firms. 
Here, we  consider an instant \emph{replacement} of firms leaving, such that the total number of firms, $N$, is  kept constant.
The new firms create links to randomly chosen partners with a small probability $q$, hence the expected number of new links is roughly $N(T)q$, where $N(T)=N-N^{\mathrm{ex}}(T)$ is the number of firms in the network after $N^{\mathrm{ex}}(T)$ firms left at time $T$.
The random choice of partners is justified by the fact that newcomers do not have complete knowledge about all established firms and their connections.

With this \emph{entry-exit dynamics} the network \emph{topology} continuously changes at time scale $T$.
We recall that it has changed already in the first intervention scenario because of firms leaving the network.
This has lead to a \emph{decreasing} number of nodes and links.
Now, instead, we have approximately a \emph{constant} number of nodes and links, but a more pronounced rewiring of the topology resulting from the combined entry and exit of firms.

Interestingly, the fact that a loss of firms is always compensated does \emph{not} imply that large cascades of firms leaving are prevented.
A random addition of new firms does not guarantee that these firms are also well integrated in the network, therefore they may leave rather soon.
So, what is the advantage of this network intervention?
It boosts the dynamics of the whole network and occasionally leads to improvements.
In fact, firms leaving may open up chances for new firms connecting to the network, and this way generating a better knowledge transfer.
Therefore, we keep the random replacement of firms leaving as \emph{one element} of our intervention scenario, and call it the \emph{network intervention}. 
Further, to ensure a continuous network dynamics, we will replace the firm with the lowest $x^{\mathrm{stat}}_{i}(T)$ even in the rare case that \emph{no} firm has decided to leave.

We emphasize that the network intervention does not imply any targeted control of firms.
Instead, it ensures that the network can constantly evolve on a time scale $T$. 
We will later test its impact by comparing our combination of network intervention and node intervention
with the previous scenario, the targeted intervention of \emph{many} firms without network intervention.

\subsection{Targeting one firm}
\label{sec:interv-strat}

According to the network intervention, firms that leave the network at time step $T$ are replaced by new firms that randomly link to existing firms with a small probability $q$.
This scenario ensures a continuing development both of the network and the knowledge stock of firms.
But occasionally it can happen that \emph{no} firm would decide to leave the network, given the conditions (A) and (B).
Then, according to the network intervention, we would replace the firm with the lowest knowledge stock $x^{\mathrm{stat}}_{i}(T)$ by a new firm that randomly connects to the network, to keep the system dynamics going. 

We now propose our \emph{node intervention}: instead of forcing the firm with the lowest knowledge stock to leave the network,
we force a \emph{specific} firm which we call \emph{firm 1}, to \emph{leave} the network, with probability $p=1$.
We choose as \emph{firm 1} a firm that \emph{initially} belongs to the \emph{core} of the network.
This firm is then replaced by a new firm 1.
In case the new firm 1 may no longer be part of the core, we will not apply our node intervention to firm 1 unless it becomes part of the core of the network, again.

This intervention approach sounds very odd, as our aim is to prevent cascades.
Firms located at the \emph{core} of the network have a higher knowledge stock and are often involved in cycles.
Therefore their impact on the cascade dynamics is expected to be even larger.
This raises two questions. Are we able to ``sacrifice'' a core firm \emph{without} enforcing a large cascade?
The answer is yes, and it will be illustrated in the small example below.
The second and most obvious question is whether such an intervention approach is indeed able to prevent large cascades of firms leaving.
The answer is yes, again, although it is quite counter-intuitive.
But it is known already from the forest fire model that small forest fires have the ability to prevent large forest fires \citep{Zinck2009}.
Therefore, it makes sense, from time to time, to remove a specific firm from the network in an \emph{ordered and controlled procedure}, to avoid situations in which large cascades can happen.

To let firm 1 leave the network implies an intervention that brings the knowledge stock of firm 1 below the threshold. We achieve our goal by simply increasing the cost $c$ of only this firm, i.e., $c\equiv c+u_{1}\delta_{1,i}$.
The Kronecker delta is $\delta_{1,i}=1$ only if $i=1$ and 0 otherwise.
The critical level of $u_{1}$ is defined by the condition $x_{1}^{\mathrm{stat}}(T)-x^{\mathrm{thr}}<0$, i.e., $u_{1}$ depends on the current network, but only needs to be chosen sufficiently large. 

\paragraph{An example. \ }

The complexity of our model results from the fact that the network of firms changes on a time scale $T$, which impacts the possible outcome of the stationary knowledge stocks of all firms.
Therefore, we have to resort to numerical investigations.
However, for the simple case of only two firms, we are able to provide analytical insights which shall demonstrate that this intervention is indeed able to drive the network dynamics to different states.

Let us assume two firms 1 and 2, where firm 1 is subject of the targeted intervention, as assumed above.
Each firm has only one directed link towards the other firm, and the dynamics of their knowledge stocks follows from Eqn.~\eqref{eq:18}:
\begin{align}
  \label{eq:2nodes.control}
    \frac{d x_1(t)}{d t}=&-\gamma\, x_1(t) + b\, x_2(t) - [c+u_{1}] x_1^{2}(t) \nonumber \\
    \frac{d x_2(t)}{d t}=&-\gamma\, x_2(t) + b\, x_1(t) - c\, x_2^{2}(t)
\end{align}
If $u_{1}=0$, the stationary knowledge stock of both firms is given as
\begin{align}
  \label{eq:1}
  x^{\mathrm{stat}}_{1}=x^{\mathrm{stat}}_{2}=\frac{b-\gamma}{c}
\end{align}
If $u_{1}\neq 0$, then instead we find for the stationary values
\begin{align}
  \label{eq:2}
  x^{\mathrm{stat}}_{1}=\mathcal{Q}\;;\quad
  x^{\mathrm{stat}}_{2}=\frac{\mathcal{Z}}{b}\;; \quad
  \mathcal{Z}=[c+u_{1}]\mathcal{Q}^{2}+\gamma\mathcal{Q}
\end{align}
where $\mathcal{Q}$ is a quite involved function of the parameters $b$, $c$, $u_{1}$ and $\gamma$, which we print in Appendix~\ref{sec:appendix-expl-stat-solut}.
From Eqn.~\eqref{eq:2} we find that (i) the values of \emph{both} $x^{\mathrm{stat}}_{1}$ and $x^{\mathrm{stat}}_{2}$ depend on the intervention $u_{1}$ and (ii) that these values are decreasing functions of $u_{1}$.
That means, for \emph{both} firms:
\begin{equation}
  \label{eq:3}
  x^{\mathrm{stat}}_{1}(u_{1}>0)<x^{\mathrm{stat}}_{1}(u_{1}<0) \;; \quad
    x^{\mathrm{stat}}_{2}(u_{1}>0)<x^{\mathrm{stat}}_{2}(u_{1}<0)
\end{equation}
The first inequality is understandable since a negative value of $u_{1}$ would \emph{decrease} the cost of firm 1, and hence increase its stationary knowledge stock.
Because the influence of the intervention $u_{1}$ propagates from firm 1 to firm 2, we also find an increased stationary knowledge stock for firm 2. 
Still, the influence of $u_{1}$  on firm 1 is always stronger:
\begin{align}\label{eq:relations_two_nodes}
	x^{\mathrm{stat}}_{1}(u_{1}>0)<x^{\mathrm{stat}}_{2}(u_{1}>0) \;; \quad 
	x^{\mathrm{stat}}_{2}(u_{1}<0)>x^{\mathrm{stat}}_{1}(u_{1}<0)
\end{align}
From Eqn.~\eqref{eq:relations_two_nodes}, we see that for a given thres\-hold $x^{\textrm{thr}}>0$, there exist a critical level $\tilde u_{1}>0$ such that
\begin{equation}
  \label{eq:4}
  x^{\mathrm{stat}}_{1}(\tilde u_{1}) < x^{\mathrm{thr}} < x^{\mathrm{stat}}_{2}(\tilde u_{1}).
\end{equation}
This is proven in a Theorem presented in the Appendix.       
That means we can always find a critical level  $\tilde u_{1}$ for the intervention such that firm 1 would \emph{leave} the network, but firm 2 would \emph{stay} in the network, even though it is impacted by the intervention applied to firm 1.

\begin{figure}[htbp]
  \centering
\includegraphics[width=.46\textwidth]{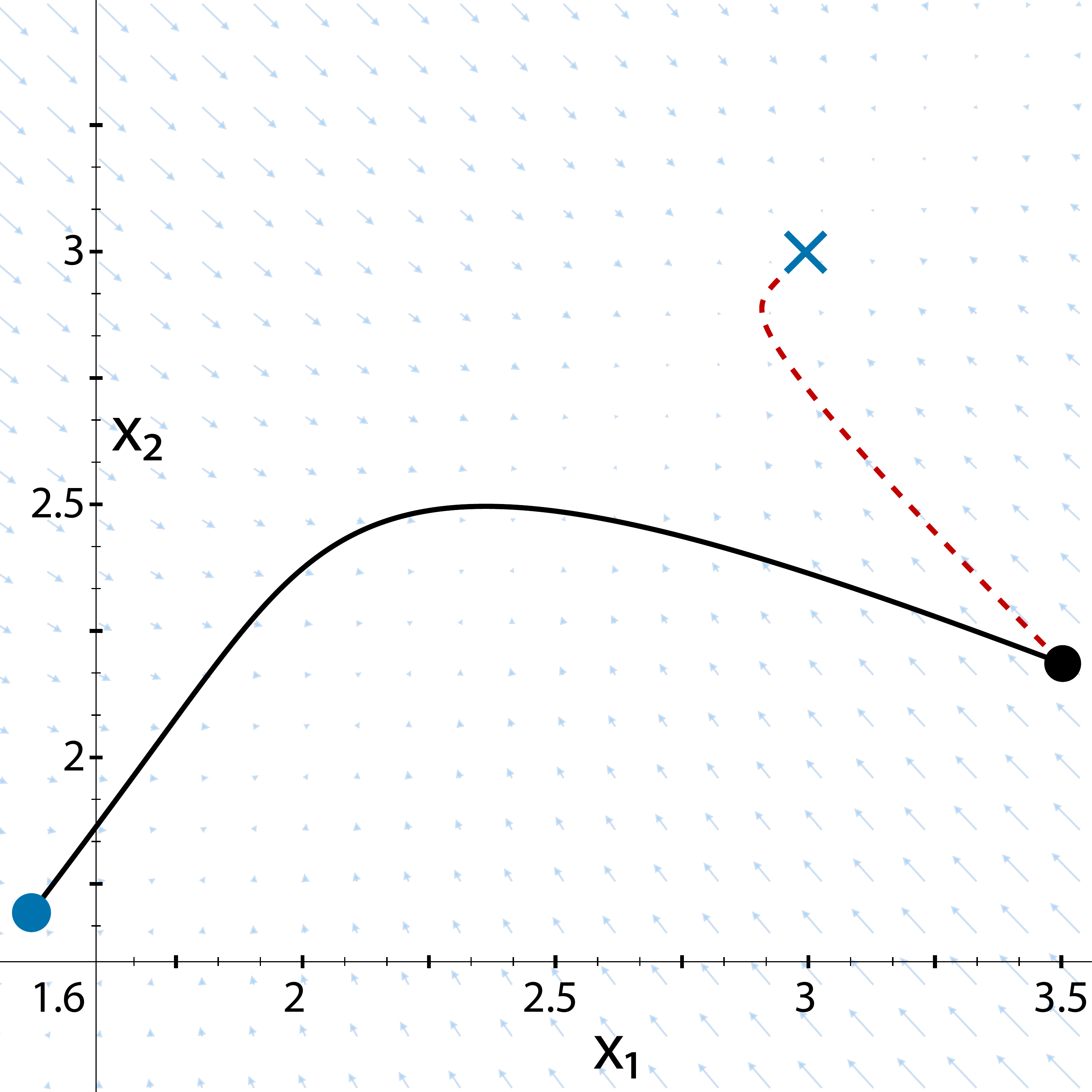}
\caption[Controlled vector field]{Phase plot illustrating the dynamics of the knowledge stocks $x_{1}(t)$, $x_{2}(t)$ of two firms, Eqn.~\eqref{eq:2nodes.control}. (red dashed line) $u_{1}\equiv 0$, (black solid line) $u_{1}=0.2$.  
  }
  \label{fig:controlfield}
\end{figure}

We illustrate the impact of the intervention on firm 1 in  Figure~\ref{fig:controlfield}.
It shows the dynamics of the knowledge stock of the two firms with and without intervention by means of two trajectories  in a so-called phase plot, $(x,y)\equiv (x_{1},x_{2})$.
Both firms initially have a knowledge stock above the threshold given by $x^{\mathrm{thr}}=1.6$.
The initial condition is marked by the black dot. 
Without any intervention, the knowledge stock of both firms would evolve along the dashed red line to the stationary solution marked by the blue cross.
The vector field shown indicates this dynamics.
However, with an intervention of firm 1, i.e., $u_{1}=0.2$, we are able to drive the dynamics such that it follows the black line and ends up in a stationary solution where $x_{1}^{\mathrm{stat}}<x^{\mathrm{thr}}$, while $x^{\mathrm{stat}}_{2}>x^{\mathrm{thr}}$, indicated by the blue dot.
This demonstrates that our intervention to ``sacrifice'' firm 1, i.e., to force it to leave, indeed works because it  will not, at the same time, causes the other firm to leave. 
Because there is no simple induction from the case of two firms to the case of $N$ firms, we will have to show the efficiency of the heuristic approach by means of computer simulations, presented in the following.

\subsection{Results of computer simulations}
\label{sec:results}

To illustrate our heuristic approach of combining network interventions and node interventions we choose a rather small network with $N=20$.
Its evolution is studied over very long time $\mathcal{T}$, to check the robustness of the system state achieved by our interventions.

\begin{figure}[htbp]
\begin{subfigure}{.32\textwidth}
    \includegraphics[width=\textwidth]{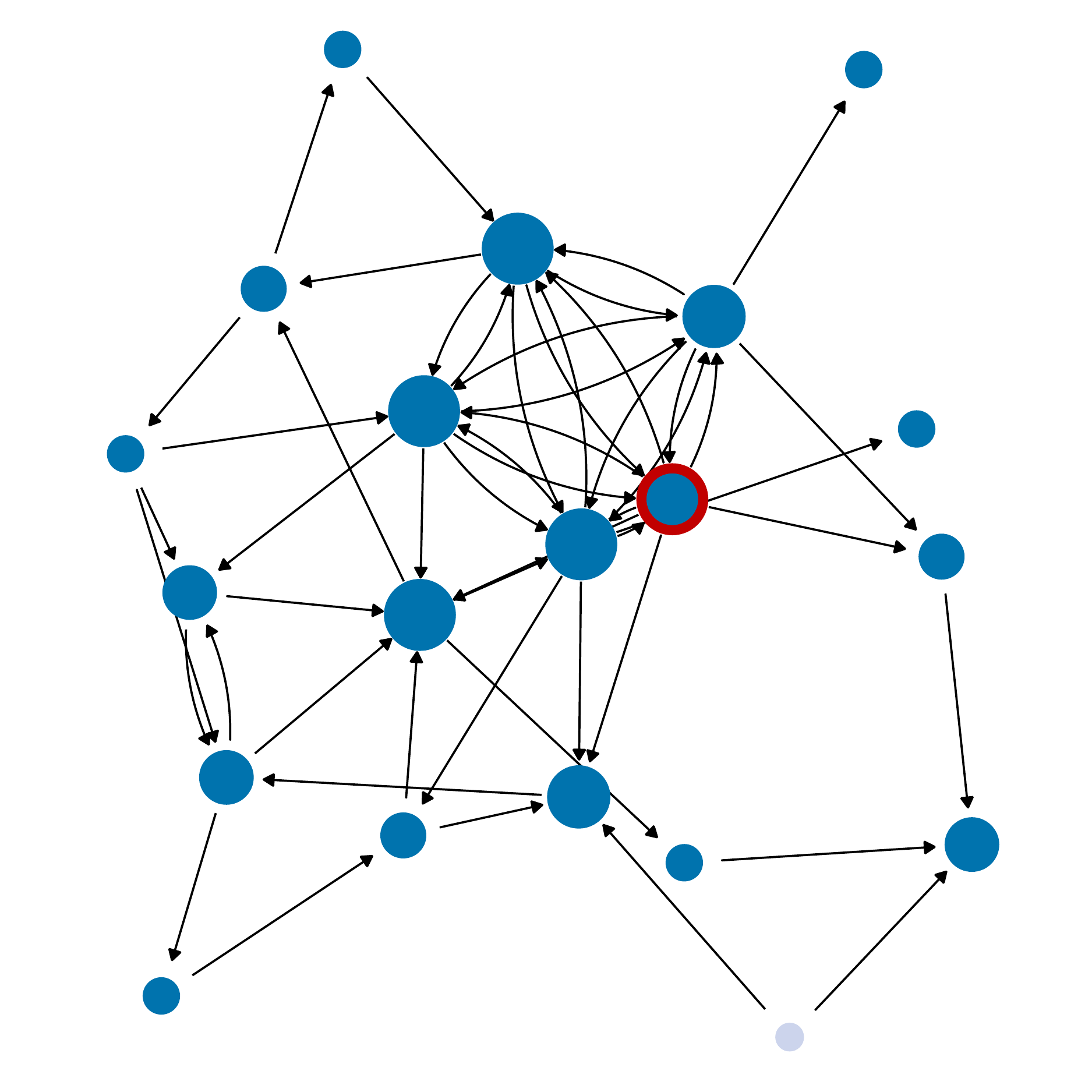}
    \caption{T=250}
\end{subfigure}
\hfill
\begin{subfigure}{.32\textwidth}
\includegraphics[width=\textwidth]{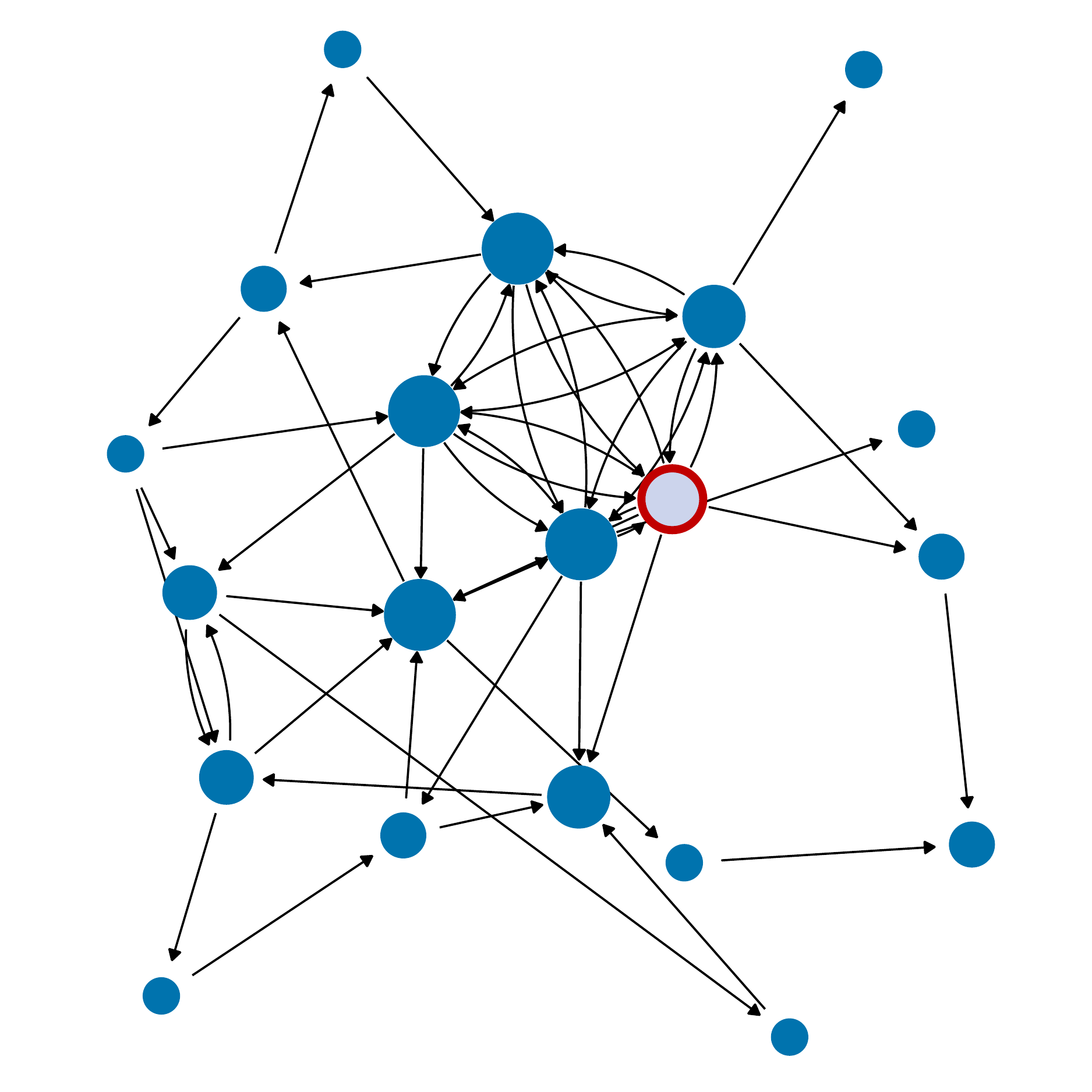}
    \caption{T=251}
\end{subfigure}
\hfill
\begin{subfigure}{.32\textwidth}
    \includegraphics[width=\textwidth]{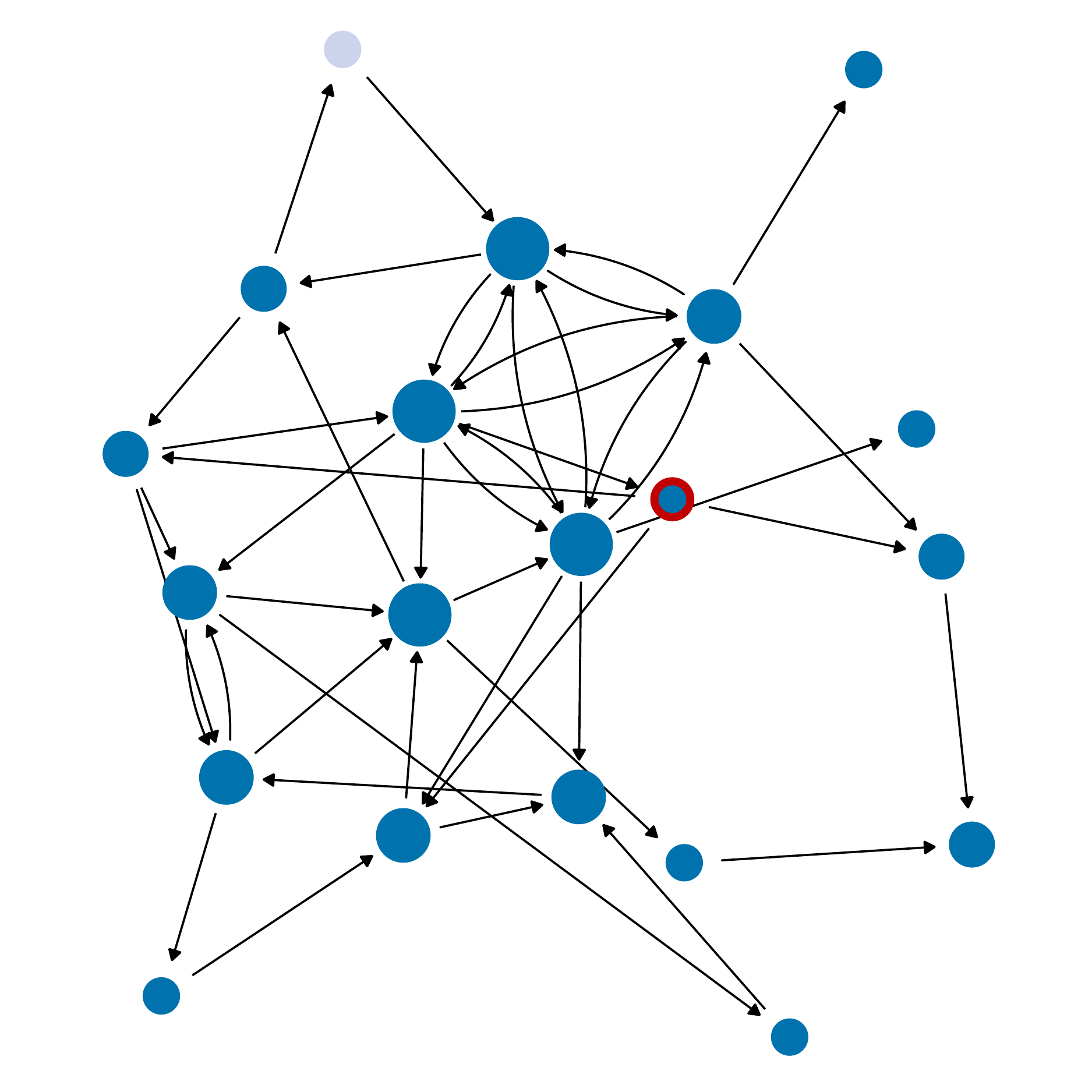}
        \caption{T=252}
\end{subfigure}
\caption{Network evolution with a targeted intervention of \emph{firm 1} (always circled). Firms with $x_{i}^{\mathrm{stat}}(0)<x^{\mathrm{thr}}$ (condition A) (marked in light blue) will leave the network.
  }
  \label{fig:fixcontrol-graph}
\end{figure}

Figure~\ref{fig:fixcontrol-graph} shows the network from one run of our simulations, at three consecutive time steps. 
For the decision of firms to leave we apply condition (A) and a leaving probability $p=1$.
The color code indicates the stationary knowledge stock of each firm.
Light blue means that firms have a value $x_{i}^{\mathrm{stat}}(T)$  lower than the threshold $x^{\mathrm{thr}}$ and thus will leave the network immediately. 
According to the network intervention, they are replaced by new firms that randomly connect to the network.
This part of our intervention scenario is applied at every time step $T$.

Until $T=250$, there were always one or more firms with a low knowledge stock that decided to leave.
At $T=251$, Firm 1, which was already \emph{initially} identified as a node from the core of the network, is targeted with an intervention for the first time.
The intervals at which such a node intervention becomes necessary fluctuate considerably.
Averaging over a long time and many simulations, we find that this intervention happened about every 11 time steps, i.e., in about 11\% of network interventions.
The control signal forces firm 1 to leave the network, and a new firm 1 enters and randomly links itself to the existing firms.
This changes the adjacency matrix and results in new stationary knowledge stocks for all firms.
The subsequent snapshot at the next time step is shown in Figure~\ref{fig:fixcontrol-graph}(c).
We note that ``sacrificing'' firm 1 has resulted in a network with an \emph{increased periphery}.

What have we gained from this node intervention?
First of all, we have ensured that the network dynamics continues, which always has the potential to also reach states of \emph{better} knowledge exchange between firms.
Secondly, we have generated a \emph{small cascade} of firms leaving.
While this seems to be unnecessary at the current point, it may become important at a later time because it may prevent a \emph{larger cascade} in the future.

\begin{figure}[htbp]
  \includegraphics[width=.45\textwidth]{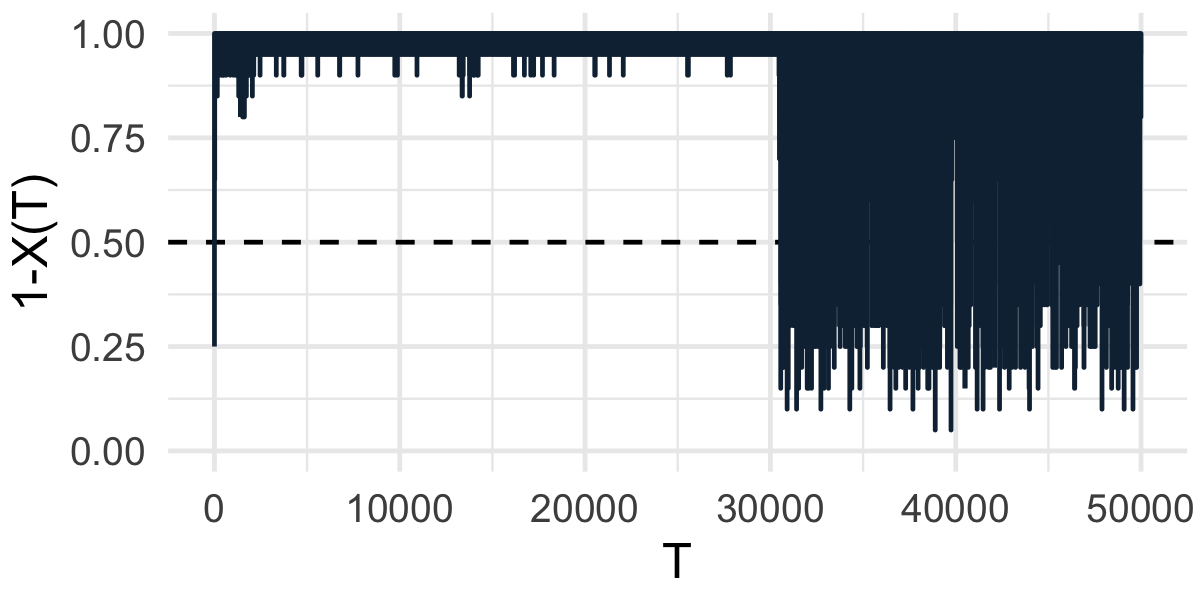}(a)
\hfill
  \includegraphics[width=.45\textwidth]{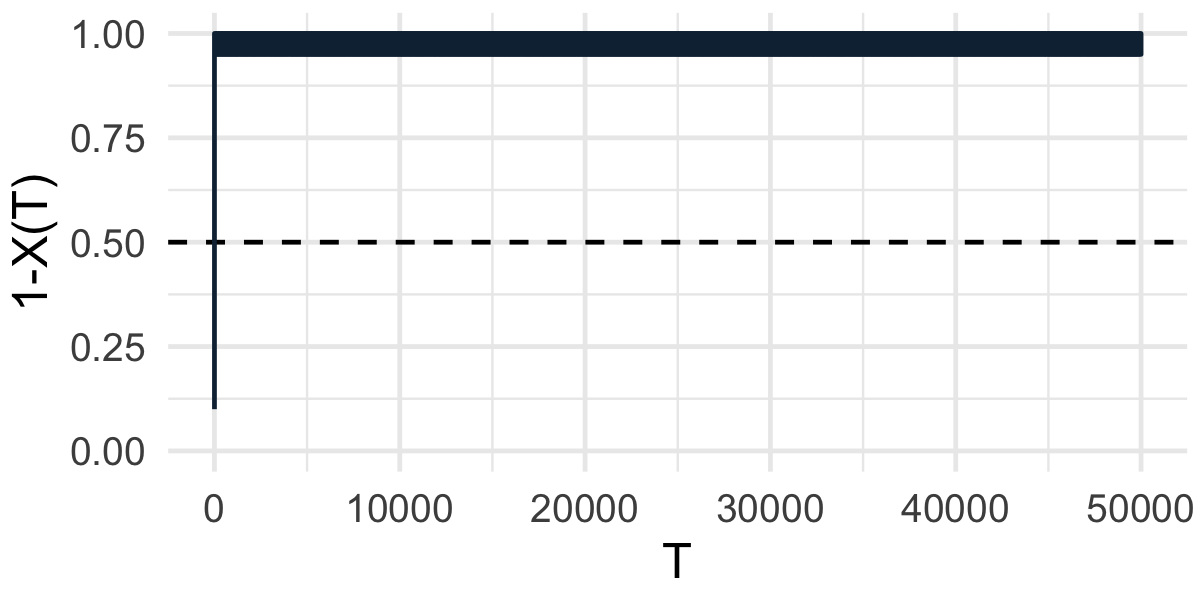}(b)
    \caption{Fraction of firms remaining in the network, $1-X(T)$, over time $T$ for different interventions: (a) network intervention only, (b) network interventions and node interventions combined. 
}
  \label{fig:fix20}
\end{figure}

To demonstrate this effect, we have plotted, from a single run, the fraction of firms remaining in the network, $1-X(T)$, over time $T$ for the cases of (a) only network intervention and (b) combined network and node intervention.
The results are shown in Figure~\ref{fig:fix20}(a,b).
We recall that \emph{without} the network intervention, we would observe a cascade of firms leaving such that the system dynamics ends rather soon (see also Figure~\ref{abm:figure1}).
\emph{With} only the network intervention, we can prolong the existence of the network considerably as Figure~\ref{fig:fix20}(a) shows.
There are always cascades of firms leaving with varying size.
But it takes a considerable long time before the network breaks down completely.
Hence, we can conclude that the \emph{network} intervention alone already allows to \emph{delay} the breakdown for a time $\mathcal{T}$.

This finding can be contrasted with Figure~\ref{fig:fix20}(b), which shows that a combination of network and node interventions is in fact able to \emph{prevent} the breakdown of the network, completely.
Even for a very long time horizon $\mathcal{T}\simeq 50\times 10^{3}$, we do not notice \emph{larger} cascades.
We remind that cascades always happen at every time step $T$, but their size is small in case of combined interventions.

Again, to account for random effects in generating the networks we have run 100 simulations to obtain values for $\hat{T}$, which is the time where the cascade has caused more than 50\% of the firms leaving. 
For the scenario with combined interventions, we always find that $\hat{T}>\mathcal{T}$, i.e., there is no breakdown observed.
But it is still interesting to note the impact of network interventions only.
There we find an average value for $\hat{T}$ of about 13.000 time steps, but a median of about 1.300 time steps only.
So, again we have a left skew distribution.
While there \emph{is} an improvement of delaying cascades in comparison to the case of no intervention, we can argue that because of the rather small median, this improvement is not large, in particular not if we compare it to the improvement from the combined interventions.

To check the robustness of our results, we have also carried out simulations with $N=250$ nodes, a time horizon $\mathcal{T}\simeq 10\times 10^{3}$ time steps, and 50 runs.
They confirm the findings discussed above.
For the combined interventions, we do not observe a breakdown for the given time horizon.
Considering network interventions only, we find that both the average and the median are about the same, so instead of a left skew distribution we have a more symmetric one.
With 7.100, their value is close to the chosen time horizon.
That means, a significant number of runs does not show large cascades during this period.
This suggests that network interventions alone become quite efficient for larger networks.

\section{Conclusions}
\label{sec:conclusions}

Interventions are seen as one important possibility to improve the state of a system.
To quantify what kind of interventions would lead to an ``improvement'' requires an understanding of both the structure and the dynamics of the target system.
Thus, the topic can hardly be discussed in an abstract manner. 
In this paper, we focus on a collaboration network of firms which exchange knowledge to increase their knowledge stock.
Such R\&D networks play a vital role in innovation economics and are therefore already studied empirically and theoretically \citep{tomasello2016,Koenig2009a,abrahamson1997social,powell2005nda}.
Hence, our investigations can build on established models for the dynamics of knowledge stock and knowledge exchange.
Less studied, so far, are possible intervention scenarios to improve knowledge exchange. 

In our paper, we have addressed one particular challenge for the R\&D network, namely the fact that firms  terminate their collaborations and leave the R\&D network if their expectations about knowledge exchange are not met.
This is a rational behavior because collaborations are costly, thus the knowledge gain has to overcome some critical threshold.
Specifically, to model the decision to leave we have extended previous works by considering not only the firm's knowledge stock, but also their knowledge growth, i.e., their expectations about the development.  

If firms decide to leave the network, this negatively impacts their collaborators and may lead to cascades of firms leaving.
With our different intervention scenarios, we want to mitigate this situation.
``Improvement'' means to either prevent such cascades, to reduce or to delay them, i.e., to increase the \emph{robustness} of the R\&D network. 
Our reference case is a scenario without any investigations, for which we can calculate the size of cascades.

In our investigations, we make use of the fact that the dynamics of our model involves two different time scales.
On the shorter time scale $t$ the knowledge stock of each firms changes, which determines whether firms stay or leave the collaboration network.
Based on their decisions, the network changes on a larger time scale $T$. 
This modeling assumption allows us to separate the two time scales.
We are interested in the systems dynamics not asymptotically, but only for a certain time horizon $\mathcal{T}$, measured in time steps $T$. 
We consider this \emph{finite} time horizon as relevant for managers or politicians to monitor the network dynamics, to plan for interventions, and to implement them.

Our intervention scenarios follow a bottom-up approach.
That means, instead of improving the situation of all firms equally by changing some global conditions, we try to identify a smaller set of firms that should be incentivized.
Such firms are called driver nodes in network controllability.
Formal approaches to identify minimal sets of driver nodes exist \citep{YZ-ACS,Cornelius2013,Jia2013f,Wang2012e,Liu2011d}.
We apply them here for the case of a non-linear dynamics of knowledge growth.

Further, we compare two different ranking schemes for driver nodes.
In one case, we choose the 20\% of firms with the highest knowledge stock, which are mostly high-degree nodes, in the other case we choose instead the 20\% of firms with the highest control contribution, which are often low-degree nodes.
Control contribution is a novel node level measure \citep{YZ-ACS} that considers the probability of a node to become a driver node and the size of the subnetwork that is controlled by this node. 

If a cascade of firms leaving is about to start, i.e., if 5\% of firms already left, 
driver nodes are incentivized to stay in the network, by lowering their costs by 10\%.
We emphasize that this is a comparably small intervention, applied to only a small subset of 20\% of all firms and only at critical times.
But this intervention is sufficient to considerably delay the emergence of a cascade of firms leaving, as the results in Figure~\ref{abm:figure2} demonstrates.
We also find that driver nodes chosen for their high control contribution are more effective in preventing cascades of firms leaving.

Our second intervention scenario uses a different approach, by combining interventions on the node and on the network level.
Network interventions are not targeted at specific firms.
Instead, replacing firms that left by new ones that randomly connect to the remaining firms allows a continuous dynamics of the network.
This takes into account that random changes \emph{can} also lead to an improvement.
In situations where not much is to lose because cascades already happen, it is certainly worth to be considered. 

This intervention alone, however, is not able to prevent cascades of firms leaving, it can only considerably delay them.
The reason comes from the fact that new firms are usually not well integrated in the network.
Therefore, to improve the situation, we also use a node intervention, but only targeted at \emph{one} driver node and only at about 10\% of all time steps. 
This firm is forced to \emph{leave} the network, i.e., instead of reducing the costs, we increase them, for only this firm, such that it is no longer attractive for the firm to stay.

While this intervention seems to be counter-intuitive, it indeed stabilizes the network such that large cascades are prevented, as Figure~\ref{fig:fix20} illustrates.
That means, the \emph{controlled removal} of one firm, chosen the right way at the right time, sustains the network of knowledge transfer \emph{if} it is combined with the network intervention.
The logic of this combined intervention is somewhat similar to the mentioned wildfire prevention \citep{Zinck2009}.
Allowing small wildfires from time to time reduces the risk of a large wildfire considerably.
Here, we demonstrate that this logic can be successfully transferred also to an economic context.

Comparing the two different intervention scenarios discussed in this paper, we point to a number of commonalities and differences.
First of all, both scenarios are successful, if we consider our minimal aim to prevent large cascades for a certain time horizon $\mathcal{T}$.
That means, we could buy considerable time to further improve the situation by, e.g., management or policy decisions.
The maximal aim, namely to prevent large cascades completely, can be also achieved in the second scenario.
But from an economic viewpoint, this might not even be desirable.
We remind on Schumpeter's idea of ``creative destruction'' as an important ingredient of innovation dynamics and economic progress \citep{metcalfe2017creative}.
From this perspective, cascades of firms leaving the collaboration network continuously test the stability, or the viability, of the knowledge transfer system.
Preventing the breakdown of an inefficient economic system might have advantages on very short time scales, but definitely hamper its long-term evolution.

A major difference of both intervention scenarios is related to costs.
In the first scenario, there is the cost of identifying and accessing the driver nodes, which is not included in our model.
But it is obvious that firms with a high knowledge stock, which are mostly nodes with a high degree, are likely more difficult to access and to influence.
It could be a bit easier, from this perspective, to access firms with a high control contribution which have a low degree.
Then, there is the cost of incentivizing the driver nodes, by lowering their costs, which has to be multiplied by the number of driver nodes.
Even if the incentives are small, they can sum up to a considerable amount.

In the second scenario, on the other hand, the intervention is not to reduce the costs of the target firm, but to increase it.
Additionally, only one firm needs to be controlled.
The network intervention, which is an important part of the second scenario, is not associated with costs for incentives.
Instead, the costs involved for the new firms to establish links to the network are covered by these firms.

Eventually, the main reason why the different intervention scenarios have any impact on the knowledge transfer between firms results from the \emph{interaction dynamics}.
Both the positive (reducing the costs) or the negative (increasing the costs) interventions \emph{not only} impact the knowledge stock of the targeted firms. 
Because of the \emph{network effects} underlying the knowledge production, these interventions propagate to other firms and steer the 
evolution of the whole system toward a desired system state.
This state is characterized by a larger robustness against cascades of firms leaving the network.

We emphasize that this way we reach a desired state that is an \emph{emerging property} of the \emph{system}, not the merit of one or a few firms.
Of course, interventions are targeted at specific firms.
But we remind that firms have only control over their \emph{outgoing} links, while their knowledge growth is entirely determined by \emph{incoming} links.
This points back to the earlier discussion about the role of \emph{indirect} reciprocity in knowledge transfer and innovation processes.
Achieving a desired outcome by means of bottom-up interventions is diametral to a greedy optimization procedure that only accepts direct improvements of the agents involved. 
Instead, the robustness of the knowledge transfer network or, the resilience of a system comprising a large number of agents in general, is the result of targeted interventions that leverage indirect reciprocity and network effects.

\small \setlength{\bibsep}{1pt}

\normalsize 
\appendix

\section{Appendix}
\label{sec:appendix1}

\subsection{Explicit stationary solutions}
\label{sec:appendix-expl-stat-solut}

Here we provide the detailed expressions for the stationary knowledge stocks of a network with only two firms 1,2.
Their dynamics was given in Eqn.~\eqref{eq:2nodes.control} and their stationary solutions in Eqn.~\eqref{eq:2}, which we repeat here:
\begin{align*}
x^{\mathrm{stat}}_{1}=\mathcal{Q}\;;\quad
  x^{\mathrm{stat}}_{2}=\frac{\mathcal{Z}}{b}\;; \quad
  \mathcal{Z}=[c+u_{1}]\mathcal{Q}^{2}+\gamma\mathcal{Q}
\end{align*}
The function $\mathcal{Q}$ reads explicitely: 
\begin{align}
  \label{eq:r2stcontrol}
\mathcal{Q}=&\frac{\sqrt[3]{k}}{3 \sqrt[3]{2} \left[-c^3-2 u_{1} c^2-u_{1}^2 c\right]}+\frac{2
              \left(\gamma c^2+\gamma u_{1} c\right)}{3 \left[-c^3-2 u_{1} c^2-u_{1}^2 c\right]} \nonumber \\
              & -\frac{\sqrt[3]{2} \left\{3 \left(-c
        \gamma^2-b c \gamma-b u_{1} \gamma\right) \left[-c^3-2 u_{1} c^2-u_{1}^2 c\right] -4 \left(\gamma c^2+\gamma u_{1}
        c\right)^2\right\}}{3 \left[-c^3-2 u_{1} c^2-u_{1}^2 c\right] \sqrt[3]{k}}
\end{align}
with $k$ given by:
\begin{align}
  \label{eq:kr1stcontrol}
  k=& -27 b^3 c^6-2 \gamma^3 c^6+9 b \gamma^2 c^6-108 b^3 u_{1} c^5-6 \gamma^3 u_{1} c^5+36 b \gamma^2 u_{1} c^5 -162 b^3 u_{1}^2 c^4-6 \gamma^3
      u_{1}^2 c^4 \nonumber \\
    & +54 b \gamma^2 u_{1}^2 c^4-108 b^3 u_{1}^3 c^3  -2 \gamma^3 u_{1}^3 c^3+36 b \gamma^2 u_{1}^3 c^3-27 b^3 u_{1}^4 c^2+9 b \gamma^2 u_{1}^4
  c^2+\sqrt{m}
\end{align}
and $m$ given by:
\begin{align}
  \label{eq:mr1stcontrol}
  m= & 4 \left\{3 \left(-c \gamma^2-b c \gamma-b u_{1} \gamma\right) \left[-c^3-2 u_{1} c^2-u_{1}^2 c\right] -4 \left(\gamma c^2+\gamma u_{1}
       c\right)^2\right\}^3 \nonumber \\
  &+\Big\{-27 b^3 c^6-2 \gamma^3 c^6+9 b \gamma^2 c^6-108 b^3 u_{1} c^5-6 \gamma^3 u_{1} c^5+36 b \gamma^2
    u_{1} c^5-162 b^3 u_{1}^2 c^4-6 \gamma^3 u_{1}^2 c^4  \nonumber \\
  & +54 b \gamma^2 u_{1}^2 c^4-108 b^3 u_{1}^3 c^3-2 \gamma^3 u_{1}^3 c^3+36 b \gamma^2 u_{1}^3
    c^3-27 b^3 u_{1}^4 c^2+9 b \gamma^2 u_{1}^4 c^2\Big\}^2
\end{align}

\subsection{Theorem for controlling two firms}
\label{sec:appendix-2}

Equation~\eqref{eq:relations_two_nodes} states that, given a thres\-hold $x^{\mathrm{thr}}>0$, we can always find
a critical value of the intervention $\tilde u_{1}$ such that the stationary knowledge stock of firm 2 (without intervention)
is above the threshold, while the knowledge stock of firm 1 (targeted by the intervention) is below the threshold.
We can formalise this statement in a simple theorem.

\begin{defn}
  Let $x^{\mathrm{thr}}>0$ be a threshold and $x_{i}^{\mathrm{stat}}(u_{1})$ be the stationary
  knowledge stocks of firms $i\in\{1,2,\}$ with a control signal $u_{}\in\mathbb R$ applied to firm 1.
  Then, as long as $x^{\mathrm{thr}}\leq x_{2}^{\mathrm{stat}}(0)$, there is a
  value $\tilde u_{}\in\mathbb R$ such that:
  \begin{equation}
    \label{eq:description}
    x_{1}^{\mathrm{stat}}(\tilde u_{})\leq x^{\mathrm{thr}}\leq x_{2}^{\mathrm{stat}}(\tilde u_{})\,.
  \end{equation}
\end{defn}
In the following, we provide a sketch of its proof.
  There are two possible cases:
  \begin{align*}
    \label{eq:proof1}
    x^{\mathrm{thr}}= x_{1}^{\mathrm{stat}}(0)= x_{2}^{\mathrm{stat}}(0)\\
x^{\mathrm{thr}}< x_{1}^{\mathrm{stat}}(0)= x_{2}^{\mathrm{stat}}(0) \end{align*}
  The first case is trivial.
  Choosing $\tilde u=0$ is the  solution.
  For the other case, we use the expressions given in Eqns.~\eqref{eq:2} and  
  \eqref{eq:r2stcontrol}.
  With $i\in\{1,2\}$, we can see $x^{\mathrm{stat}}_{i}(u_{})\in\left(0, x^{\mathrm{stat}}_{i}(0)\right]$ as monotonously decreasing and continuous functions of
  $u_{}\in[0,\infty)$.
  Hence, they are also bijective functions within this interval.
  According to Eqn. \eqref{eq:relations_two_nodes}, then, there exist a positive $\tilde u$ such that Eqn. \eqref{eq:description} is
  satisfied.

\end{document}